\documentclass[sigconf]{acmart}
\AtBeginDocument{%
  }
\usepackage{paralist}
\usepackage[most]{tcolorbox}
\newcommand{\eatme}[1]{ }
\usepackage{threeparttable}
\usepackage{listings}
\usepackage{amsmath}
\usepackage{graphicx}
\usepackage{caption}
\usepackage{subcaption}
\usepackage{algorithm}
\usepackage{algpseudocode}

\setcopyright{none}
\acmISBN{}
\acmDOI{}
\acmYear{}

\begin{document}

\title{Benchmarking Dimensionality Reduction Techniques for \\ Spatial Transcriptomics}

\author{%
Md Ishtyaq Mahmud\textsuperscript{1},
Veena Kochat\textsuperscript{3},
Suresh Satpati\textsuperscript{3},
Jagan Mohan Reddy Dwarampudi\textsuperscript{1},\\
Kunal Rai\textsuperscript{3},
Tania Banerjee\textsuperscript{1,2}}
\thanks{\textsuperscript{1}\;Department of Electrical and Computer Engineering, University of Houston, Houston, TX, USA}
\thanks{\textsuperscript{2}\;Department of Information Science Technology, University of Houston, Sugar Land, TX, USA}
\thanks{\textsuperscript{3}\;Department of Genomic Medicine and MDACC Epigenomics Therapy Initiative (METI), MD Anderson Cancer Center, Houston, TX, USA}
\thanks{Emails: krai@mdanderson.org, tbanerjee@uh.edu}



\renewcommand{\shortauthors}{Mahmud et al.}

\begin{abstract}
We introduce a unified framework for evaluating dimensionality reduction techniques in spatial transcriptomics beyond standard PCA approaches. We benchmark six methods—PCA, NMF, autoencoder, VAE, and two hybrid embeddings—on a cholangiocarcinoma Xenium dataset, systematically varying latent dimensions ($k$=5-40) and clustering resolutions ($\rho$=0.1-1.2). Each configuration is evaluated using complementary metrics including reconstruction error, explained variance, cluster cohesion, and two novel biologically-motivated measures: Cluster Marker Coherence (CMC) and Marker Exclusion Rate (MER). Our results demonstrate distinct performance profiles: PCA provides a fast baseline, NMF maximizes marker enrichment, VAE balances reconstruction and interpretability, while autoencoders occupy a middle ground. We provide systematic hyperparameter selection using Pareto optimal analysis and demonstrate how MER-guided reassignment improves biological fidelity across all methods, with CMC scores improving by up to 12\% on average. This framework enables principled selection of dimensionality reduction methods tailored to specific spatial transcriptomics analyses.

\end{abstract}



\keywords{Spatial transcriptomics, Dimensionality Reduction, 
Principal component analysis (PCA), Nonnegative matrix factorization (NMF), Autoencoder (AE), Variational autoencoder (VAE), Cluster Marker Coherence (CMC), Marker Exclusion Rate (MER), MER-guided refinement, Cell reassignment, Cholagiocarcinoma, Tissue microarray}


\maketitle

\section{Introduction}
\begin{figure}[htbp]
  \centering
  \includegraphics[width=0.6\columnwidth]{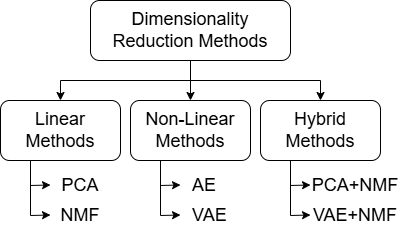}
  \caption{An overview of the dimensionality reduction methods evaluated in this work, grouped into linear (PCA, NMF), non-linear (AE, VAE), and hybrid (PCA+NMF, VAE+NMF) approaches. The hybrid approach uses concatenated embeddings from two methods.}
  \label{fig:dim-methods-overview}
\end{figure}

Dimensionality reduction is a cornerstone of modern spatial transcriptomics pipelines, transforming each cell’s high-dimensional gene-expression profile into a compact embedding that both denoises the data and highlights biologically meaningful variation.  In practice, the de facto standard for dimensionaliy reduction is Principal Component Analysis (PCA), a linear projection that maximizes variance along orthogonal axes.  However, PCA may miss nonlinear or parts-based structure.  In this work, we systematically benchmark two additional classes of dimensionality methods (shown in Figure~\ref{fig:dim-methods-overview}:

\begin{itemize}
  \item \textit{Linear methods}: PCA and Non-negative Matrix Factorization (NMF). NMF constrains both factors and loadings to be nonnegative, yielding additive, parts-based representations.
  \item \textit{Deep nonlinear methods}: Autoencoders (AE) and Variational Autoencoders (VAE), which learn flexible encoder–decoder networks that can capture complex manifolds in gene-express-ion space.
  \item \textit{Hybrid methods}: Concatenated embeddings, namely PCA + NMF and VAE + NMF, that combine complementary linear and nonlinear features.
\end{itemize}

To evaluate how each embedding shapes downstream clustering, we employ a suite of metrics:

\begin{itemize}
  \item \emph{Reconstruction fidelity:} Mean squared error (MSE) and explained variance of the embedding.
  \item \emph{Clustering quality:} Silhouette score and Davies–Bouldin Index (DBI).
  \item \emph{Biological coherence:} two novel metrics — Cluster Marker Coherence (CMC), the fraction of cells in each cluster expressing its marker genes, and Marker Exclusion Rate (MER), the fraction of cells that would express another cluster’s markers more strongly.
  \item \emph{Gene-set enrichment:} average enrichment of known marker-gene sets per cluster.
\end{itemize}

Because clustering solely on low-dimensional embeddings can misassign cells whose original transcriptomes express another cluster's markers, we introduce a lightweight post-processing step. Any cell with higher aggregate marker expression in a different cluster is reassigned accordingly, guided by MER. We then perform a before-versus-after analysis to quantify how this refinement improves marker coherence and downstream clustering metrics. Note that we exclude spatially-augmented algorithms from this analysis, reserving them for future work.

Our contributions are threefold:
\begin{enumerate}
  \item To our knowledge, this represents the first systematic comparison of PCA, NMF, AE, VAE, and hybrid embeddings applied to the same spatial transcriptomics data. By controlling for dataset variability, we can directly assess how mathematical differences in dimensionality reduction translate into biological differences in the resulting cluster assignments and marker gene expression patterns.
  \item We introduce two novel clustering metrics, Cluster Marker Coherence (CMC) and Marker Expression Ratio (MER), that are interpretable and biologically motivated, and designed to assess how well clustering results align with marker gene expression patterns.
  \item We present a simple yet novel MER-guided cell reassignment algorithm that substantially enhances marker coherence through post-processing refinement of initial clustering results.
  
\end{enumerate}

The remainder of the paper is organized as follows.  Section~\ref{sec:related} reviews related work.  Section~\ref{sec:theory} presents a unified mathematical framework for dimensionality reduction methods.  Section~\ref{sec:method} describes our datasets, preprocessing, implementation of the dimensionality reduction methods and evaluation metrics. Section~\ref{sec:expt} reports our experimental results and discusses practical implications, and finally, Section~\ref{sec:conclude} concludes the paper.  

\eatme{
\begin{figure}[htbp]
  \centering
  \includegraphics[width=\columnwidth]{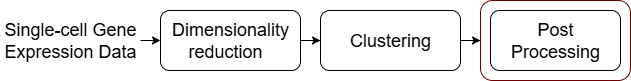}
  \caption{Overview of our end‐to‐end workflow for spatial transcriptomics analysis. Starting from raw single‐cell gene expression data, we apply dimensionality reduction (PCA, NMF, AE, VAE, or hybrids), followed by clustering (e.g.\ Leiden). Finally, we introduce a post‐processing step based on Marker Exclusion Rate (MER) to refine cluster assignments and improve biological coherence.}
  \label{fig:workflow-pipeline}
\end{figure}

\begin{figure}[htbp]
  \centering
  \includegraphics[width=\columnwidth]{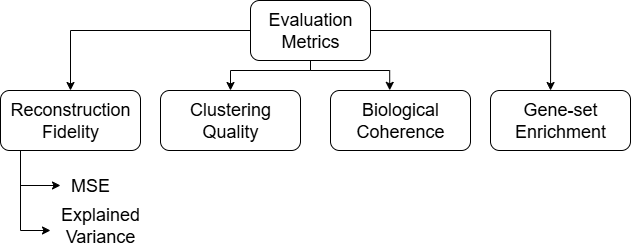}
  \caption{Overview of the metrics that are used in the paper.}
  \label{fig:workflow-pipeline}
\end{figure}
}

\eatme{
\begin{table*}[ht]
\centering
\caption{Conceptual comparison of dimensionality reduction methods}
\label{tab:dr_comparison}
\begin{tabular}{|p{2.2cm}|p{4cm}|p{5cm}|p{5cm}|}
\hline
\textbf{Method} & \textbf{Objective Function} & \textbf{Constraints / Regularization} & \textbf{Interpretability} \\
\hline
PCA &
$\min_{W} \|X - XWW^\top\|_F^2$ &
$W^\top W = I$ (orthogonality) &
Low: linear, dense components \\
\hline
NMF &
$\min_{Z, W} \|X - ZW^\top\|_F^2$ &
$Z, W \geq 0$ (non-negativity) &
High: additive parts-based factors \\
\hline
Autoencoder (AE) &
$\min_{\theta, \phi} \|X - g_\phi(f_\theta(X))\|_F^2$ &
Neural weights; optionally sparse/l1/l2 regularization &
Medium: nonlinear, flexible but opaque \\
\hline
Variational AE (VAE) &
Maximize ELBO: $\mathbb{E}_{q(z|x)}[\log p(x|z)] - \text{KL}(q(z|x)\|p(z))$ &
KL divergence (prior regularization) &
Medium: smoother latent space, less entangled \\
\hline
Graph AE (GAE) &
$\min \|X - g_\phi(f_\theta(X, A))\|$ &
Spatial graph $A$; Laplacian smoothing &
High (spatial): encodes local structure \\
\hline
\end{tabular}
\end{table*}
}

\eatme{
\section{Related Work (Draft - rewrite)}
Dimensionality reduction (DR) is a critical step in the analysis of high-dimensional single-cell and spatial transcriptomics data, enabling downstream tasks such as clustering, visualization, and identification of tissue-specific cell niches. Numerous DR methods have been proposed and adapted for biological data, yet few systematic evaluations have been conducted in the context of spatial omics.

\subsection{DR Methods in Transcriptomics and Spatial Data}

Classical linear methods such as principal component analysis (PCA) remain widely used in single-cell pipelines due to their simplicity and speed. PCA is often employed in tools like Seurat \cite{satija2015spatial} and Scanpy \cite{wolf2018scanpy} as a first-pass dimensionality reduction before neighborhood graph construction and clustering. However, PCA may not capture nonlinear biological variation or spatial coherence.

Non-negative matrix factorization (NMF) has shown utility in uncovering additive gene programs and modules \cite{brunet2004nmf}, and has been applied in spatial transcriptomics to identify interpretable spatial domains \cite{cheng2021spatialsort}. Due to its parts-based decomposition, NMF is particularly appealing for interpretability, though it lacks flexibility for modeling nonlinear interactions.

Deep learning-based models, such as autoencoders (AEs) and variational autoencoders (VAEs), offer expressive nonlinear mappings that can capture complex manifolds in gene expression data. Several recent works have adapted AEs and VAEs to single-cell analysis \cite{eraslan2019autoencoders, lopez2018deep}, but their role in spatial transcriptomics remains underexplored. Challenges such as interpretability, overfitting, and disentanglement persist in these models without strong inductive biases or regularization.

Graph-based methods such as graph autoencoders (GAEs) explicitly incorporate spatial adjacency information, which is particularly relevant in spatial omics. Recent tools like SpaGCN \cite{hu2021spagcn} and STAGATE \cite{dong2022deciphering} leverage graph neural networks to integrate spatial and expression information for domain discovery. While effective, these approaches are often task-specific and rarely compared systematically across datasets or to simpler DR methods.

\subsection{Benchmarking and Comparative Studies}

Recent benchmarking efforts in single-cell analysis, such as \cite{duo2020systematic, luecken2022benchmarking}, have evaluated DR techniques in terms of clustering and visualization, but these largely ignore spatial features and the specific interpretability demands of spatial transcriptomics. In the spatial omics context, studies like \cite{zhao2021spatialdim} have begun to assess method performance, but comparisons often focus on clustering outcomes without theoretical framing or interpretability analysis.

\textcolor{blue}{Seurat was presented by \cite{satija2015spatial}, a groundbreaking pipeline that allows for the spatial reconstruction of gene expression in single cells. This reconstruction was indirect and relied on scRNA-seq that was mapped back to spatial patterns, though it did not use spatial expression data like Visium or xenium.
The researcher \cite{brunet2004nmf} came up with non-negative matrix factorization (NMF) as a way to find metagenes and hidden biological structure in gene expression data. However, the study was only about bulk transcriptomics and overlook at spatial organization or cellular resolution, which are two of the biggest problems in spatial omics. STAGATE is a graph attention autoencoder that was developed by another group of researchers \cite{dong2022deciphering}. It was designed specifically for spatial transcriptomics and shown excellent spatial domain recovery on both Visium and Slide-seq data. On the other hand, their model is tuned for spatial segmentation, and it was not benchmarked in comparison to simpler DR models like as PCA, NMF, or AE.  They proposed \cite{eraslan2019autoencoders} a deep count autoencoder for the purpose of denoising data obtained from single-cell RNA sequencing, but it does not contain any spatial modeling components and was never tested in the context of spatial transcriptomics. Another group of researchers \cite{hu2021spagcn} presented SpaGCN, a graph convolutional network that integrates geographical coordinates, gene expression, and histology in order to identify spatial domains. Despite the fact that it displayed higher performance in domain identification, the majority of its attention was directed towards segmentation tasks. Furthermore, it did not investigate the reconstruction quality or the theoretical interpretability of the learnt embeddings. A deep generative model for probabilistic modeling of single-cell RNA-seq data called scVI was presented \cite{lopez2018deep}. This model is built on variational autoencoders and developed for single-cell, non-spatial contexts, but it does not contain spatial priors or neighborhood structure. Scanpy is a scalable toolbox that was developed by them \cite{wolf2018scanpy} for the purpose of analyzing gene expression in single cells. The original implementation of Scanpy was not spatially aware and did not integrate spatial coordinates or morphology. Despite its widespread use for dimensionality reduction, clustering, and visualization, Scanpy's implementation was not spatially aware. They \cite{gronbech2020scvae} studied variational autoencoders (VAEs) using biologically limited latent spaces to improve single-cell gene expression data representation learning. They employed deep probabilistic modeling to capture transcriptome variability's generative structure in scVAE. The model showed potential in denoising, imputation, and displaying complex gene expression patterns. No tissue coordinates or local neighborhood structure were used in the development and validation of the work, which used non-spatial scRNA-seq datasets. The study did not evaluate spatial coherence, spatial domain preservation, or benchmark performance across spatial transcriptomics platforms like Visium, Slide-seq, and Xenium, restricting its applicability to spatial omics activities.
}

Our work aims to fill this gap by providing a comprehensive and methodologically unified comparison of DR techniques—PCA, NMF, AE, VAE, and GAE—in the context of spatial transcriptomics. We emphasize both empirical benchmarking and theoretical interpretability, including spatial coherence, biological relevance, and reconstruction quality. To our knowledge, this is the first evaluation that integrates these dimensions across multiple spatial transcriptomics platforms.

References:~\cite{satija2015spatial}, \cite{wolf2018scanpy}, \cite{brunet2004nmf}, \cite{eraslan2019autoencoders}, \cite{lopez2018deep}, \cite{hu2021spagcn}, \cite{dong2022deciphering}, \cite{luecken2022benchmarking}
}
\vspace{-0.4cm}
\section{Related Work\label{sec:related}}

Dimensionality reduction is a fundamental step in the analysis of high-dimensional single-cell and spatial transcriptomics data, underpinning downstream tasks such as clustering, visualization, and tissue-domain discovery. While many dimensionality reduction methods exist, few studies have systematically compared their performance and interpretability in a spatial-omics setting.

\paragraph*{Classical and Matrix-Factorization Methods:}

PCA remains the de facto choice in single-cell pipelines (e.g., Seurat~\cite{satija2015spatial}, Scanpy~\cite{wolf2018scanpy}) due to its analytical tractability and speed. However, its linear projections can miss nonlinear biological variation or spatial context.

NMF uncovers additive, parts-based gene programs~\cite{brunet2004nmf} and has been applied to spatial data for interpretable domain discovery~\cite{Townes2023}. Its nonnegativity constraint yields intuitive gene signatures, but it cannot model nonlinear interactions.

\paragraph*{Deep Learning Approaches:}

AEs and VAEs provide flexible, nonlinear mappings that can capture complex manifolds in single-cell expression~\cite{eraslan2019autoencoders, lopez2018deep}. Despite their promise, these models have rarely been benchmarked in spatial transcriptomics, and challenges remain around implementation, overfitting, interpretability, and alignment with spatial structure.


\paragraph*{Benchmarking Studies}

Several recent studies have compared dimensionality reduction techniques on single-cell RNA-seq data, evaluating clustering tightness, global structure preservation, and visualization clarity~\cite{duo2020systematic, sun2024}, including surveys of existing techniques such as PCA, t-SNE, and scVI~\cite{lopez2018deep}. However, these analyses typically do not provide in-depth comparisons by applying multiple techniques to identical datasets with systematic parameter exploration. Our work fills this gap by providing a unified mathematical framework alongside comprehensive empirical evaluation of linear (PCA, NMF), nonlinear (AE, VAE), and hybrid (PCA+NMF, VAE+NMF) approaches on the same spatial transcriptomics dataset. We explicitly measure embedding quality, introduce novel metrics for marker coherence (CMC) and marker exclusion rate (MER), and analyze how different dimensionality reduction methods affect cluster composition and biological interpretation within tissue microarray samples.

\section{Theoretical Framework and Interpretability\label{sec:theory}}

In this section we lay out a unified view of the four dimensionality‐reduction techniques under study, PCA, NMF, AE and VAE, in terms of their core optimization objectives, and briefly discuss how these formulations influence interpretability in the spatial‐transcriptomics setting: Let
\[
  X \in \mathbb{R}^{n \times d}
\]
be our normalized cell–by–gene expression matrix (with \(n\) cells and \(d\) genes), and let
\[
  Z \in \mathbb{R}^{n \times k}
\]
denote its low‐dimensional embedding (\(k \ll d\)).  We then have:

\begin{compactenum}
    \item \textit{Principal Component Analysis (PCA):} PCA solves the minimization problem
    \[
    \min_{Z, W} \|X - ZW^\top\|_F^2 \quad \text{subject to } W^\top W = I,
    \]
    where $W \in \mathbb{R}^{d \times k}$ is an orthonormal basis and $Z = XW$ is the projection of $X$ onto the $k$-dimensional subspace~\cite{jolliffe2016principal, Zhao2021}.

    \item \textit{Non-negative Matrix Factorization (NMF):} NMF introduces non-negativity constraints to produce parts-based, additive components~\cite{Lee1999}:
    \[
    \min_{Z \geq 0, W \geq 0} \|X - ZW^\top\|_F^2.
    \]
    This formulation often yields interpretable gene signatures and cell states by aligning with the inherently non-negative nature of expression data~\cite{brunet2004nmf}.

    \item \textit{Autoencoder (AE):} AEs are nonlinear encoder-decoder networks trained to minimize reconstruction error:
    \[
    \min_{\theta, \phi} \|X - g_\phi(f_\theta(X))\|_F^2.
    \]
    The encoder $f_\theta$ maps $X$ to a latent representation $Z$, and the decoder $g_\phi$ reconstructs $X$ from $Z$. Without explicit regularization, AEs may yield non-interpretable representations~\cite{hinton2006reducing, eraslan2019autoencoders}.

    \item \textit{Variational Autoencoder (VAE):} VAEs model a probabilistic latent space via an encoder $q_\theta(z|x)$ and decoder $p_\phi(x|z)$, optimized using the Evidence Lower Bound (ELBO):
    \[
    \mathbb{E}_{q_\theta(z|x)}[\log p_\phi(x|z)] - \text{KL}(q_\theta(z|x) \| p(z)).
    \]
    The KL divergence term enforces a prior on the latent distribution (usually Gaussian), promoting disentangled and regularized latent spaces~\cite{kingma2022autoencodingvariationalbayes, Hu2024}.
\eatme{
    \item \textbf{Graph Autoencoder (GAE):} GAEs extend AEs to graph-structured data by incorporating a cell-cell adjacency matrix $A$, often derived from spatial proximity or gene expression similarity:
    \[
    Z = f_\theta(X, A), \quad \hat{X} = g_\phi(Z).
    \]
    This structure allows GAE to preserve local tissue architecture, making it particularly well-suited to spatial omics applications~\cite{Kipf2016VariationalGA, kipf2017semisupervisedclassificationgraphconvolutional}.
    }
\end{compactenum}
In practice, choosing among these methods requires trading off \emph{reconstruction accuracy}, \emph{clustering quality}, and \emph{biological interpretability}, these are the very axes along which we evaluate each technique in the remainder of this paper.

\eatme{
\subsection{PCA vs Linear Autoencoders}

Principal Component Analysis (PCA) is a classical linear dimensionality reduction technique that projects the input data $X \in \mathbb{R}^{n \times d}$ onto a $k$-dimensional subspace that maximizes variance. The PCA projection matrix $W_{\text{PCA}} \in \mathbb{R}^{d \times k}$ is composed of the top-$k$ eigenvectors of the data covariance matrix, and the projection $Z = X W_{\text{PCA}}$ minimizes the reconstruction error under the constraint that $W^\top W = I$.

A linear autoencoder can be viewed as a generalization of PCA in which the encoder and decoder are parameterized as linear functions. Specifically, let $f_\theta(x) = Wx$ and $g_\phi(z) = W^\top z$ for $W \in \mathbb{R}^{k \times d}$, and assume the data is zero-centered. Then the autoencoder learns $W$ by minimizing the reconstruction loss $\|X - XWW^\top\|_F^2$. Unlike PCA, the linear AE is typically trained via gradient descent without requiring orthogonality or eigen decomposition. When tied weights are used, and no nonlinearities are applied, the linear AE recovers the same subspace as PCA.

\begin{lemma}\label{lemma:pca}
Let $X \in \mathbb{R}^{n \times d}$ be a zero-centered data matrix, where rows represent $n$ cells and columns represent $d$ genes in a spatial transcriptomics dataset. Consider a linear autoencoder with encoder $f_\theta(x) = Wx$, decoder $g_\phi(z) = W^\top z$, and shared weight matrix $W \in \mathbb{R}^{k \times d}$ (i.e., tied weights) trained to minimize the mean squared reconstruction error:
\[
\min_W \|X - XWW^\top\|_F^2.
\]
The optimal solution $W^*$ spans the same $k$-dimensional subspace as the top-$k$ principal components of $X$, and the minimum reconstruction error equals that of PCA: $\|X - XWW^\top\|_F^2 = \|X - X_k\|_F^2$, where $X_k$ is the rank-$k$ truncated SVD approximation of $X$~\cite{baldi1989linear}. See Appendix~A for a proof sketch.
\end{lemma}

This equivalence establishes PCA as a baseline for linear dimensionality reduction in spatial transcriptomics, which we compare against nonlinear methods like VAEs and GAEs.
}

\eatme{
\begin{tcolorbox}[colback=blue!5!white,colframe=blue!75!black,title=Theoretical Insight: PCA Sets a Lower Bound for Linear AEs]
In the case of linear encoders and decoders with tied weights, an autoencoder cannot outperform PCA in terms of reconstruction error. PCA provides the optimal rank-$k$ projection under the Frobenius norm. Nonlinear AEs may exceed PCA performance, but at the cost of reduced interpretability and the need for regularization.
\end{tcolorbox}

\begin{tcolorbox}[colback=blue!5!white,colframe=blue!75!black,title=Insight: PCA as a Linear AE Baseline]
PCA provides the optimal rank-$k$ projection for linear autoencoders with tied weights, setting a baseline for reconstruction error in spatial transcriptomics~\cite{baldi1989linear}.
\end{tcolorbox}

\subsection{Variational Autoencoders and Information Regularization}

The Variational Autoencoder (VAE) extends the traditional autoencoder by introducing a probabilistic framework for learning latent representations. Instead of learning a single point $z = f_\theta(x)$, the encoder approximates a posterior distribution $q_\theta(z|x)$ over latent variables, and the decoder defines a conditional likelihood $p_\phi(x|z)$. The VAE is trained to maximize the Evidence Lower Bound (ELBO) on the marginal log-likelihood of the data:
\[
\mathcal{L}_{\text{VAE}}(x) = \mathbb{E}_{q_\theta(z|x)}[\log p_\phi(x|z)] - \text{KL}(q_\theta(z|x) \| p(z)),
\]
where $p(z)$ is typically chosen to be a standard multivariate normal prior. The first term encourages accurate reconstruction, while the second term regularizes the latent space by pushing $q_\theta(z|x)$ to remain close to the prior $p(z)$. This KL divergence term plays a crucial role in controlling the information flow from $x$ to $z$, and governs the tradeoff between expressivity and generalization in the learned embedding.

\begin{lemma}\label{lemma:vae}
Let $q(z|x)$ be the variational posterior, $p(z)$ the prior, and $q(x, z) = p_{\text{data}}(x) q(z|x)$ the joint distribution for a VAE applied to a spatial transcriptomics dataset. The mutual information between $x$ and $z$ satisfies:
\[
I_q(x; z) \leq \mathbb{E}_{p_{\text{data}}(x)} \left[ \text{KL}(q(z|x) \| p(z)) \right].
\]
See Appendix~B for a proof sketch~\cite{Zhao_Song_Ermon_2019}.
\end{lemma}

This bound highlights the role of the KL divergence in regularizing VAE latent spaces, which we benchmark against PCA and other methods for cell type clustering.
}

\eatme{
\begin{tcolorbox}[colback=green!5!white,colframe=green!75!black,title=Theoretical Insight: KL Divergence Bounds Mutual Information in VAEs]
The KL term in the VAE loss upper-bounds the mutual information between input data and latent variables. This introduces an inherent tradeoff: minimizing KL divergence regularizes the latent space (improving generalization), but may suppress meaningful signal if over-penalized. This explains why VAEs often show smoother but less discriminative embeddings compared to AEs.
\end{tcolorbox}

\begin{tcolorbox}[colback=green!5!white,colframe=green!75!black,title=Insight: VAE Latent Space Regularization]
The KL divergence in VAEs bounds mutual information, promoting smooth but less discriminative embeddings for spatial transcriptomics tasks~\cite{Zhao_Song_Ermon_2019}.
\end{tcolorbox}

\paragraph{Choice of VAE Baseline over Spatially-Augmented Variants}
We evaluate a standard VAE rather than spatially-augmented variants like SpaCAE~\cite{Hu2024} for two reasons. First, SpaCAE is optimized for spot-based platforms like 10x Visium with grid-like spatial structures, which is less suitable for the single-cell resolution of Xenium data. Second, a vanilla VAE isolates the effects of probabilistic regularization (Lemma~\ref{lemma:vae}), allowing spatial coherence gains in GAE (Lemma ~\ref{lemma:gae}) and our novel spatially-regularized NMF (Lemma~\ref{lemma:snmf}) to be attributed to their graph-based structures rather than combined spatial and contrastive objectives.
}

\eatme{
\subsection{Interpretability in NMF and Spatial Coherence in GAE}

NMF decomposes a non-negative gene expression matrix into interpretable cell contributions and gene modules, with identifiability guaranteed under separability (Lemma~\ref{lemma:nmf}). We benchmark NMF against PCA, VAEs, and GAEs to evaluate its effectiveness in capturing biologically meaningful patterns in spatial transcriptomics, such as co-expressed gene clusters or spatial niches.

Graph Autoencoders (GAEs), in contrast, incorporate prior knowledge of spatial or functional relationships among cells. Given a graph structure represented by an adjacency matrix $A$, the encoder learns node embeddings $Z = f_\theta(X, A)$ by aggregating features from neighboring nodes, while the decoder reconstructs the original feature matrix or the adjacency itself. The objective typically takes the form:
\[
\min_{\theta, \phi} \|X - g_\phi(f_\theta(X, A))\|_F^2,
\]
where $f_\theta$ is often implemented using graph convolutional layers. The inclusion of graph-based smoothing leads to spatially coherent embeddings that preserve tissue topology, making GAEs well-suited for tasks such as niche discovery in spatial transcriptomics.

\begin{lemma}\label{lemma:nmf}
Let $X \in \mathbb{R}^{n \times d}$ be a non-negative data matrix, where rows represent $n$ cells and columns represent $d$ genes in a spatial transcriptomics dataset. Suppose $X \approx WH$, with non-negative matrices $W \in \mathbb{R}^{n \times k}$ (cell contributions) and $H \in \mathbb{R}^{d \times k}$ (gene modules). If $X$ satisfies the separability condition—i.e., for each column of $H$, there exists a row in $X$ that is a scaled version of that column—then the factorization is identifiable up to permutation and scaling of the columns of $W$ and rows of $H$~\cite{Arora2012}. See Appendix~C for a proof sketch.
\end{lemma}
}
\eatme{
\begin{tcolorbox}[colback=orange!5!white,colframe=orange!75!black,title=Theoretical Insight: NMF Enables Interpretable Factorizations Under Mild Conditions]
Unlike PCA and AEs, Non-negative Matrix Factorization (NMF) can yield unique and interpretable decompositions under a simple geometric condition known as separability. In biological contexts, this aligns with the assumption that some cells or genes strongly represent distinct functional modules (e.g., anchor genes). This supports NMF’s use for discovering gene programs or interpretable latent clusters.
\end{tcolorbox}

\begin{tcolorbox}[colback=orange!5!white,colframe=orange!75!black,title=Insight: NMF’s Interpretable Gene Modules]
NMF yields unique, interpretable gene modules under separability, ideal for discovering functional clusters in spatial transcriptomics~\cite{Arora2012}.
\end{tcolorbox}
}

\eatme{
\begin{lemma}\label{lemma:gae}
Let $A$ be the adjacency matrix of a spatial cell graph in a spatial transcriptomics dataset, and $L = D - A$ the unnormalized graph Laplacian, where $D$ is the degree matrix. Suppose a Graph Autoencoder’s encoder applies a low-pass spectral filter $g_\theta(L)$ to input features $X \in \mathbb{R}^{n \times d}$ (gene expression for $n$ cells, $d$ genes). Then the learned embedding $Z = g_\theta(L)X$ satisfies:
\[
\text{Smoothness}(Z) = \text{Tr}(Z^\top L Z) \leq \text{Tr}(X^\top L X),
\]
i.e., the embeddings are smoother over the graph than the input features~\cite{kipf2017semisupervisedclassificationgraphconvolutional}. See Appendix~D for a proof sketch.
\end{lemma}

\eatme{
\begin{tcolorbox}[colback=purple!5!white,colframe=purple!75!black,title=Theoretical Insight: GAE Encourages Spatial Smoothness via Spectral Filtering]
Graph Autoencoders (GAEs) operate on spatial graphs, and their encoders inherently promote smooth latent embeddings across neighboring cells. This is formalized by a decrease in the Laplacian smoothness functional, which implies that GAE embeddings respect tissue topology. This property supports GAE’s observed performance in niche discovery and spatial clustering tasks.
\end{tcolorbox}
}

\begin{tcolorbox}[colback=purple!5!white,colframe=purple!75!black,title=Insight: GAE’s Spatial Smoothness]
GAEs produce smooth embeddings over spatial cell graphs, enhancing niche discovery and clustering in spatial transcriptomics~\cite{kipf2017semisupervisedclassificationgraphconvolutional}.
\end{tcolorbox}

\subsection{Spatially-Regularized NMF}
To enhance NMF for spatial transcriptomics, we propose a novel spatially-regularized NMF that combines non-negativity with spatial smoothness, ensuring interpretable and spatially coherent cell embeddings.

\begin{lemma}\label{lemma:snmf}
Let $X \in \mathbb{R}^{n \times d}$ be a non-negative data matrix, where rows represent $n$ cells and columns represent $d$ genes in a spatial transcriptomics dataset, and let $A$ be the adjacency matrix of a spatial cell graph with unnormalized graph Laplacian $L = D - A$. Suppose $X \approx WH$, with non-negative matrices $W \in \mathbb{R}^{n \times k}$ (cell contributions) and $H \in \mathbb{R}^{d \times k}$ (gene modules), is a Non-negative Matrix Factorization (NMF) optimized with a spatial smoothness constraint:
\[
\min_{W,H \geq 0} \|X - WH\|_F^2 + \lambda \text{Tr}(W^\top L W),
\]
where $\lambda > 0$ is a regularization parameter. Then the optimal $W^*$ produces cell embeddings that are smoother over the spatial graph than the input features, i.e., $\text{Tr}(W^{*\top} L W^*) \leq \text{Tr}(X^\top L X)$, and the factorization remains identifiable under the separability condition.
\end{lemma}

This novel method combines the interpretability of NMF (Lemma~\ref{lemma:nmf}) with the spatial coherence of GAEs (Lemma~\ref{lemma:gae}), and we benchmark it against PCA, VAEs, NMF, and GAEs for spatial clustering and niche discovery.

\begin{tcolorbox}[colback=red!5!white,colframe=red!75!black,title=Insight: Spatially-Regularized NMF]
Our novel spatially-regularized NMF combines interpretable gene modules with spatial smoothness, enhancing cell embeddings for spatial transcriptomics tasks (Lemma~\ref{lemma:snmf}).
\end{tcolorbox}

\eatme{
\subsection{Interpretability and Identifiability Constraints}

Interpretability and identifiability are essential criteria in spatial transcriptomics, as the biological significance of latent components directly influences downstream insights.

\begin{compactenum}  
    \item \textbf{PCA} produces orthogonal components that maximize variance, but these components are dense linear combinations of all input genes, often complicating biological interpretation.

    \item \textbf{NMF} enforces additive and sparse representations, aligning well with biological modules (e.g., co-expressed gene clusters). Under mild assumptions such as separability, NMF is provably identifiable up to permutation and scaling. This means the factorization $V = WH$ is unique up to permutation and scaling under certain conditions, such as, separability. Separability assumes that each basis vector in $W$ corresponds to a unique data point or a distinct feature that is not a combination of others. Under these conditions, the factors $W$ and $H$ can be recovered uniquely, except for trivial ambiguities like permuting the order of components (e.g., swapping columns of $ W $ and rows of $ H $) or scaling them (e.g., multiplying a column of $ W $ by a constant and dividing the corresponding row of $ H $ by the same constant). This identifiability property is a theoretical strength of NMF compared to other matrix factorization methods like PCA, which may not guarantee uniqueness without additional constraints.

    \item \textbf{AE} models can capture complex, nonlinear manifolds but often yield entangled latent representations unless explicitly regularized, for example, via sparsity constraints or a contractive penalty. Here, sparsity constraints encourage the latent representations to be sparse (i.e., many latent units are close to zero), which can help disentangle representations by limiting the number of active features. Contractive penalty refers to contractive autoencoders (CAEs), which add a regularization term to the loss function proportional to the Frobenius norm of the encoder’s Jacobian ($ \|\partial f(x)/\partial x\|_F^2 $). This penalty encourages the encoder to be less sensitive to small input perturbations, promoting local smoothness and robustness in the latent space. In either case, the learned features are typically not directly interpretable.

    \item \textbf{VAE} imposes structure on the latent space via the Kullback-Leibler (KL) divergence term, which can lead to smoother and more disentangled representations. This term encourages the learned latent distribution (typically parameterized as a Gaussian) to be close to a prior distribution (often a standard normal, $ \mathcal{N}(0, I) $). KL divergence regularization can smoothen latent areas by matching close points to similar data points in the input space. When paired with $\beta$-VAEs, KL term weight can be changed to promote disentanglement. However, VAEs are not inherently identifiable (latent variables may not uniquely match data variation factors), and interpretation of latent dimensions often remains challenging without additional constraints. For example, the latent dimensions may not immediately correspond to human-interpretable features (e.g., "color" or "shape") unless the model enforces interpretability by sparsity, factorized priors, or explicit supervision. 

    \item \textbf{GAE} incorporates spatial context by encoding graph relationships among cells, which can improve the interpretability of latent factors in tissue architecture-aware tasks such as niche discovery and spatial clustering. The encoder learns a low-dimensional representation of each node that reflects both its intrinsic features and its connections, while the decoder reconstructs the graph structure or node features. By incorporating spatial relationships, the latent factors learned by GAEs can correspond to biologically meaningful patterns, such as cell types, tissue niches, or spatial domains. This makes the latent representations more interpretable for tasks where spatial organization matters, like identifying microenvironmental niches or clustering cells based on their spatial context.
\end{compactenum}
}
\eatme{
\subsection{Implications for Spatial Transcriptomics}

Dimensionality reduction methods for spatial omics must preserve both biological interpretability and spatial coherence. PCA and NMF provide interpretable linear projections but lack the flexibility to model nonlinear structure. AE and VAE offer expressive latent spaces but require careful regularization to maintain interpretability. GAE stands out by incorporating spatial adjacency, allowing it to preserve spatial relationships that are critical in tissue architecture analysis. Our benchmarking (Section~\ref{sec:results}) evaluates these methods through this theoretical lens to inform practical model selection.
}
}
\section{Methodology\label{sec:method}}
In this section, we detail our end‐to‐end analytical workflow for benchmarking dimensionality‐reduction techniques on spatial transcriptomics data. We begin by describing the Xenium dataset and preprocessing steps (Section~\ref{sec:pp}), then outline the architectures and hyperparameters for each method (Section~\ref{sec:dr}). Next, we explain our hyperparameter‐sweep strategy and selection criteria (Section 4.3), followed by the quantitative and biological metrics employed to evaluate performance (Section 4.4). Finally, we summarize how these reduced embeddings are applied to downstream clustering and visualization tasks (Section 4.5).
\subsection{Dataset and Preprocessing\label{sec:pp}}

We used Xenium spatial transcriptomics on TMA cores from \textit{N=25} cholangiocarcinoma patients (total of \textit{M=40} cores). The assay consisted of 480 target panel genes, and $\approx$ $212070$ cells. Each core yielded high‐resolution transcript counts. All samples were anony-mized TMAs obtained from the MD Anderson Cancer Center, representing intrahepatic cholangiocarcinoma resections. The following data preprocessing steps were performed first:
\subsubsection{Quality control (QC) and gene filtering.}  
From the raw Xenium count matrix, we filtered out genes detected in fewer than three cells and removed cells with fewer than 200 detected genes (including negative‐control probes). Doublet detection (cell barcodes whose transcript profiles indicated they originated from two neighboring cells merging into a single ROI) was performed using Scrublet, and following Wolock et al.~\cite{WOLOCK2019281}, we labeled any cell with a Scrublet doublet score > $0.2$ as a potential doublet. The dataset was then cleared of all such highlighted ROIs, leaving only high-confidence single-cell profiles for further examination. After QC, the dataset comprised $191,125$ spatially localized cells across all cores.  
\subsubsection{Normalization and transformation.}  
Transcript counts were normalized per cell to a total of $10,000$ counts, followed by $\log_{e}(x+~1)$ transformation.  
    
    
\subsubsection{Batch and artifact correction.}  
No explicit batch‐effect or spa-tial‐artifact correction was performed, as QC metrics indicated minimal technical variation across cores.

\subsection{Dimensionality Reduction Methods~\label{sec:dr}}
We detail the model architectures and associated hyperparameters for each dimensionality reduction method below.

\subsubsection{Principal Component Analysis (PCA)}  
We applied Scanpy’s PCA (via \texttt{sc.tl.pca}) to the log‐normalized count matrix, extracting the top \(k\) principal components using the ARPACK solver for robust performance on high‐dimensional data and a fixed \texttt{random\_state} for reproducibility.  Denoting the cell‐by‐component scores by \(Z\in\mathbb{R}^{n\times k}\) and the gene‐by‐component loadings by \(V\in\mathbb{R}^{p\times k}\), we reconstruct the expression matrix as
\[
  \widehat{X} \;=\; Z\,V^\top,
\]
where each row \(Z_i\) contains the \(k\) component scores for cell \(i\), and each column of \(V\) is one of the top \(k\) eigenvectors of the gene–gene covariance matrix.  
\eatme{
\begin{lstlisting}[caption={PCA dimensionality reduction in Python}, label={lst:pca_code}]
def run_pca_reduction():
    sc.tl.pca(adata,
              n_comps=n_components,
              svd_solver='arpack',
              random_state=random_state)
    X_reconstructed = adata.obsm['X_pca'] @ adata.varm['PCs'].T
\end{lstlisting}
}
\subsubsection{Non‐negative Matrix Factorization (NMF)}  
We applied scikit‐learn’s \texttt{NMF} to the nonnegative log‐normalized count matrix \(X\in\mathbb{R}_{+}^{n\times p}\), sweeping the factorization rank \(k\) over \(\{5,10,\dots,40\}\).  We used the coordinate‐descent solver with nonnegative‐double‐SVD initialization (\texttt{init='nndsvda'}), a maximum of $400$ iterations, and a fixed \texttt{random\_state} for reproducibility.  The model approximates  
\[
  X \;\approx\; W\,H,
\]
where \(W\in\mathbb{R}_{+}^{n\times k}\) holds the cell‐by‐component scores and \(H\in\mathbb{R}_{+}^{k\times p}\) the component‐by‐gene loadings.  Reconstructed expression is simply  
\[
  \widehat{X} = W\,H.
\]
Nonnegative double SVD (NNDSVD) initializes \(W\) and \(H\) via a rank-\(k\) truncated SVD, \(X \approx U\Sigma V^\top\), by taking the positive parts of \(U\Sigma^{1/2}\) and \(\Sigma^{1/2}V^\top\).  This sparse, data‐driven initialization accelerates convergence and often improves reconstruction fidelity.  

\subsubsection{Autoencoder (AE)}  
We implemented a fully–connected autoencoder in PyTorch that projects the \(p\)-dimensional input into a \(d\)-dimensional latent space via an encoder of width  
\[
p \;\to\; 1024 \;\to\; 512 \;\to\; 256 \;\to\; 128 \;\to\; 64 \;\to\; d,
\]  
and reconstructs back through a symmetric decoder  
\[
d \;\to\; 64 \;\to\; 128 \;\to\; 256 \;\to\; 512 \;\to\; 1024 \;\to\; p.
\]  
Each hidden layer applies a LeakyReLU activation with negative slope 0.1~\cite{Maas2013} followed by dropout at rate 0.1~\cite{Srivastava2014} to encourage robustness and sparsity.  The decoder concludes with a Softplus output~\cite{Goodfellow-et-al-2016} to guarantee nonnegative reconstructions.  Formally, for an input vector \(x\in\mathbb{R}^p\):  
\[
z = f_{\mathrm{enc}}(x), 
\quad 
\widehat{x} = f_{\mathrm{dec}}(z),
\]  
where \(f_{\mathrm{enc}}\) and \(f_{\mathrm{dec}}\) denote the encoder and decoder mappings, respectively.  We trained the model for up to 100 epochs using the Adam optimizer (\(\mathrm{lr}=10^{-4}\), \(\mathrm{weight\_decay}=10^{-5}\)) with early stopping (patience = 5) on the validation mean‐squared error. After convergence, we extracted the cell‐by‐latent embedding \(Z\in\mathbb{R}^{n\times d}\) and the reconstructed expression matrix \(\widehat X\in\mathbb{R}^{n\times p}\) from the decoder’s output.

\subsubsection{Variational Autoencoder (VAE)}  
We implemented a custom variational autoencoder in PyTorch whose encoder maps the \(p\)-dimensional input through successive linear layers of size
\[
p \;\to\; 1024 \;\to\; 512 \;\to\; 256 \;\to\; 128 \;\to\; 64,
\]
each followed by a LeakyReLU activation~\cite{Maas2013} (slope=0.1) and dropout (rate = 0.1).  Two parallel linear heads then produce the latent mean \(\mu(x)\in\mathbb{R}^d\) and log-variance \(\log\sigma^2(x)\in\mathbb{R}^d\) for a $d$-dimensional latent space.  We sample latent codes via a modified reparameterization trick,
\[
  z = \mu(x) \;+\; \epsilon \odot \sigma(x), 
  \quad \epsilon \sim \mathcal{N}(0,\,0.1^2\,I).
\]
The decoder mirrors the encoder in reverse,
\[
d \;\to\; 64 \;\to\; 128 \;\to\; 256 \;\to\; 512 \;\to\; 1024 \;\to\; p,
\]
applying LeakyReLU and dropout at each hidden layer and a final linear output layer.  All linear weights are initialized with Xavier normal~\cite{pmlr-v9-glorot10a}, and biases set to 0.01, to accelerate convergence.

\paragraph{Reparameterization}
\[
  z = \mu + \sigma \odot (\epsilon\times 0.1),
  \quad \epsilon \sim \mathcal{N}(0,I),
\]
which injects scaled noise to stabilize early training.

\paragraph{Initialization}
All \(\texttt{nn.Linear}\) layers use
\[
  W \sim \mathcal{N}\bigl(0,\tfrac{1}{\sqrt{\text{fan\_in}}}\bigr), 
  \quad b = 0.01,
\]
via PyTorch’s \texttt{xavier\_normal\_} and constant bias seeding.
 
\paragraph{VAE Training Procedure}  
We trained each VAE for up to 100 epochs on minibatches of size 512 using the Adam optimizer (\(\mathrm{lr}=1\times10^{-3}\), \(\mathrm{weight\_decay}=1\times10^{-5}\)).  At each iteration, we computed the loss
\[
\mathcal{L} = \mathrm{MSE}(x,\hat x) \;+\;\beta\,D_{\mathrm{KL}}\bigl(q(z\mid x)\,\|\,p(z)\bigr),
\]
with \(\beta=0.2\), and applied gradient clipping (max‐norm = 1.0) before updating parameters.  We evaluated validation loss after each epoch and performed early stopping if no improvement of at least 0.01 occurred for 5 successive epochs.  After training, we ran the full dataset through the model in inference mode to extract the posterior means \(\mu\) as the latent embeddings and to reconstruct \(\hat X\); we inverted any preprocessing on \(\hat X\) and clipped to nonnegative values.  

\paragraph{Kullback–Leibler (KL) Divergence}  
The KL divergence term in the loss function between two distributions \(q\) and \(p\) is defined as  
\[
  D_{\mathrm{KL}}(q\|p)
  = \int q(z)\log\!\frac{q(z)}{p(z)}\,dz,
\]
which in the discrete case becomes
\(\sum_i q(i)\log\frac{q(i)}{p(i)}\).  In VAEs, this term regularizes the approximate posterior \(q_\phi(z\!\mid\!x)\) toward the prior \(p(z)=\mathcal{N}(0,I)\), yielding smooth, structured latent spaces~\cite{kingma2022autoencodingvariationalbayes}.


\eatme{
\subsection{Interpretability and Identifiability}

Interpretability and identifiability are critical in spatial transcriptomics to ensure biologically meaningful latent representations.

\begin{compactenum}
    \item \textbf{PCA}: Produces orthogonal components maximizing variance, but dense gene combinations hinder biological interpretation~\cite{baldi1989linear}.
    \item \textbf{NMF}: Yields sparse, additive gene modules, interpretable as co-expressed clusters, and identifiable under separability~\cite{Arora2012}.
    \item \textbf{AE}: Captures nonlinear patterns but produces entangled representations unless regularized (e.g., via sparsity)~\cite{Rifai2011}.
    \item \textbf{VAE}: Promotes smooth, disentangled embeddings via KL divergence, but lacks inherent identifiability~\cite{Zhao_Song_Ermon_2019}.
\end{compactenum}
}

\subsection{Hyperparameter‐Sweep Strategy and Selection Criteria}
For each dimensionality‐reduction method, we performed a two‐dimensional grid search over latent dimensionality 
\[
k \in \{5, 10, 15, 20, 25, 30, 35, 40\}
\]
and Leiden clustering resolution 
\[
\rho \in \{0.1, 0.2, 0.3, 0.4, 0.5, 0.6, 0.7, 0.8, 0.9, 1.0, 1.2\}.
\]
At each \((k,\rho)\) combination we computed a suite of performance metrics (defined in Section 4.4):
\begin{compactenum}
    
  \item Clustering quality: Silhouette coefficient and Davies–Bouldin index, both calculated on the low-dimensional embeddings.  
  \item Biological coherence: Unweighted and size‐weighted Cluster Marker Coherence (CMC) scores per cluster.
  \item Reconstruction fidelity: Mean‐squared error and fraction of explained variance.  
  \item Biological enrichment: –\(\log_{10}\) adjusted p‐value from gene‐set enrichment analysis of cluster marker sets.  
\end{compactenum}
We then identified the Pareto‐optimal \((k,\rho)\) pairs for each method by balancing cluster cohesion (high silhouette, low DB index) against biological fidelity (high marker‐fraction and enrichment) and data reconstruction.  The selected hyperparameters maximize marker coherence while maintaining tight, well‐separated clusters.

\subsection{Evaluation Metrics}
For each \((k,\rho)\) combination (Section 4.3), we computed the following metrics on the low‐dimensional embeddings \(Z\) and reconstructed data \(\widehat X\):

\subsubsection{Reconstruction Fidelity}  
We measure global reconstruction error by the mean‐squared error  
\[
  \mathrm{MSE}
  = \frac{1}{n\,p}\sum_{i=1}^n \sum_{j=1}^p \bigl(X_{ij} - \widehat X_{ij}\bigr)^2,
\]
where \(n\) is the number of cells and \(p\) the number of genes~\cite{Bishop2006}.  We also report explained variance  
\[
  \mathrm{ExplainedVar}
  = 1 - \frac{\sum_{i,j}(X_{ij} - \widehat X_{ij})^2}{\sum_{i,j}(X_{ij} - \bar X)^2},
\]
with \(\bar X\) the grand mean of \(X\)~\cite{jolliffe2016principal}.

\subsubsection{Clustering Quality}  
We apply Leiden clustering at resolution \(\rho\) on \(Z\) and compute the Silhouette coefficient  
\[
  s(i) = \frac{b(i) - a(i)}{\max\{a(i),b(i)\}},\quad
  \mathrm{Silhouette} = \frac{1}{n}\sum_{i=1}^n s(i),
\]
where \(a(i)\) is the average intra‐cluster distance and \(b(i)\) the lowest average inter‐cluster distance~\cite{Rousseeuw1987}. Thus, a high silhouette score implies tight, cohesive clusters and clear gaps between them. We also compute the Davies–Bouldin index (DBI), which averages the worst‐case ratio of within‐cluster scatter to between‐cluster separation. Silhouette scores range from \(-1\) to \(+1\), with higher values indicating more compact, well-separated clusters, whereas Davies–Bouldin index values run from 0 upward, with lower values signaling tighter, more distinct clustering. Importantly, the silhouette metric tends to favor solutions with many small clusters (since small, tight clusters achieve high average separation), while the Davies–Bouldin index penalizes overly fragmented clustering by measuring increased within‐cluster scatter relative to between‐cluster distances.

\subsubsection{Cluster Marker Coherence (CMC)}  
We introduce the Cluster Marker Coherence (CMC) score to quantify how faithfully each cluster captures its marker genes.  For cluster \(c\) with marker set \(\mathcal{M}_c\), we define  
\[
  F_c = \frac{1}{n_c}\sum_{i\in c}
    \frac{1}{|\mathcal{M}_c|}\sum_{g\in\mathcal{M}_c}\mathbf{1}\{x_{i,g}>0\},
\]
the fraction of markers detected per cell, averaged over the \(n_c\) cells in \(c\).  We then report both the unweighted mean  
\(\displaystyle \mathrm{CMC} = \frac{1}{C}\sum_{c=1}^C F_c\)  
and the size‐weighted mean  
\(\displaystyle \mathrm{CMC}_w = \frac{1}{n}\sum_{c=1}^C n_c\,F_c\).  
A high CMC indicates strong biological coherence, making it a useful complement to geometric clustering metrics.

\eatme{
\subsubsection{Cluster Marker Recovery Fraction (CMRF)}  
To quantify how well each clustering recovers known marker genes, we define the cluster marker recovery fraction (CMRF) as the average, over all $C$ clusters, of the fraction of true marker genes that appear among the top‐$n$ ranked genes for that cluster.  Formally, let $\mathcal{M}_c$ be the known marker set for cluster $c$, and let $\widehat{\mathcal{M}}_c$ be the inferred top‐$n$ genes for the same cluster.  Then
\[
  \mathrm{CMRF}
  = \frac{1}{C}\sum_{c=1}^{C}
    \frac{\lvert \mathcal{M}_c \,\cap\, \widehat{\mathcal{M}}_c\rvert}
         {\lvert \mathcal{M}_c\rvert}\,.
\]
A CMRF of 1.0 indicates perfect recovery of all canonical markers in every cluster, whereas 0 indicates no overlap.  In our experiments we set $n = |\mathcal{M}_c|$, i.e.\ we compare exactly as many inferred genes as there are true markers, but one can choose other values of $n$ as needed.
}

\subsubsection{Enrichment Score}  
We perform gene‐set enrichment analysis on each cluster’s marker gene set using GSEA~\cite{Subramanian2005}.  The enrichment score (ES) is the maximum deviation of a running‐sum statistic that increases for genes in the set \(\mathcal{M}_c\) and decreases otherwise:  
\[
  \mathrm{ES}
  = \max_{1\le i\le p}\Biggl|\sum_{j=1}^i\bigl(\tfrac{1}{|\mathcal{M}_c|}\mathbf{1}_{j\in\mathcal{M}_c}
   -\tfrac{1}{p-|\mathcal{M}_c|}\mathbf{1}_{j\notin\mathcal{M}_c}\bigr)\Biggr|.
\]
Significance is assessed via comparison to a null distribution of phenotype‐based permutations.

\subsubsection{Marker Exclusion Rate (MER)}  
To quantify how often cells end up in clusters whose marker genes they do not express, or feebly express, we define the \emph{Marker Exclusion Rate} (MER) as follows.  Let \(X\in\mathbb{R}^{n\times p}\) be the expression matrix and \(\mathcal{M}_c\subset\{1,\dots,p\}\) the marker‐gene indices for cluster \(c\).  For each cell \(i\), we compute its best‐matching cluster  
\[
  \hat c_i \;=\;\arg\max_{c}\;\sum_{g\in\mathcal{M}_c}X_{i,g},
\]
and flag \(i\) as \emph{marker‐positive} if \(\sum_{g\in\mathcal{M}_{\hat c_i}}X_{i,g}>0\).  Denote by \(I^+\) the set of all marker‐positive cells, i.e.\ those cells that express at least one marker gene.  Then we define the global MER as
\[
  \mathrm{MER}
  \;=\;
  \frac{1}{n}
  \sum_{i\in I^+}
  \mathbf{1}\bigl(c_i \neq \hat c_i\bigr),
\]
where \(c_i\) is the Leiden‐predicted label for cell \(i\).  Equivalently, MER is the fraction of \emph{all} cells whose assigned cluster’s marker set they fail to express.  We also report the number of cells with zero  expressed marker:
\[ MZC = n - |I^+| \]

A low MER (near 0) indicates that almost every cell resides in a cluster whose markers it actually expresses, making MER a reliable error metric for downstream cell‐type annotation.  

\medskip
Together, these metrics provide a comprehensive assessment of reconstruction accuracy, embedding compactness and separation, and biological validity.

\subsection{MER-guided Post Processing}
Algorithm~\ref{alg:mer} implements a simple post-processing step that reassigns cells whose assigned cluster does not maximize their marker-gene expression. After computing, for each cell, the total expression of each cluster’s marker set, it reassigns the cell only if (1) the alternative cluster yields strictly higher marker expression by at least a small threshold  $\epsilon$, and (2) that best cluster actually has at least one expressed marker in that cell. This procedure preserves the overall geometric cohesion established by Leiden on the reduced embedding (so silhouette scores change minimally) while reducing the number of cells whose cluster labels conflict with their own marker-gene profiles.
\begin{algorithm}
\caption{MER-guided Cluster Post-Processing}
\label{alg:mer}
\begin{algorithmic}[1]  
  \Require Embedding $Z\in\mathbb{R}^{n\times k}$, cluster labels $c_i$, marker sets $\{\mathcal{M}_c\}$
  \Ensure New labels $c_i^{\mathrm{new}}$
  \State Build expression matrix $X\in\mathbb{R}^{n\times p}$ (layer or $X$).
  \State Precompute marker-gene indices for each cluster $c$: 
           $\mathcal{I}_c = \{g : g\in\mathcal{M}_c\}$.
  \For{$i \gets 1$ to $n$}
    \State \Comment{score each cluster by total marker expression in cell $i$}
    \ForAll{clusters $c$}
      \State $s_c \gets \sum_{g\in \mathcal{I}_c} X_{i,g}$
    \EndFor
    \State $\hat{c} \gets \arg\max_c\,s_c$
    \State $\Delta \gets s_{\hat{c}} - s_{c_i}$  \Comment{improvement over original}
    \If{$\Delta > \varepsilon$ and $s_{\hat{c}} > 0$}
      \State $c_i^{\mathrm{new}} \gets \hat{c}$
    \Else
      \State $c_i^{\mathrm{new}} \gets c_i$
    \EndIf
  \EndFor
  \State \Return $\{c_i^{\mathrm{new}}\}$
\end{algorithmic}
\end{algorithm}

\subsection{Downstream Analysis}
We leveraged each low-dimensional embedding \(Z\in\mathbb{R}^{n\times d}\) for three main tasks:

\paragraph{Cell-type Clustering}  
We built a 15-nearest-neighbor graph on \(Z\) (a standard choice in transcriptomic analyses after testing within a plausible range of 10–30 neighbors with minimal effect on clustering) and ran Leiden clustering~\cite{Traag2019} (with varying resolutions) to define putative cell‐type clusters, forming the basis of our spatial cell‐type atlas.

\paragraph{Visualization}  
We visualized cells directly on their spatial coordinates within the tissue microarray sample, coloring by Leiden cluster assignments to compare the spatial organization and separation of cell types across different dimensionality reduction methods.

\subsection{Implementation Details}

All experiments were conducted in Python (v3.10) using Scanpy (v1.11.2) for data preprocessing and PCA, scikit‐learn (v1.7.0) for NMF, and PyTorch (v2.0.1) for AE and VAE implementations. Models were trained on a workstation with an NVIDIA Tesla V100-PCIE-32GB GPU. Runtimes and memory overhead for representative runs of each dimensionality reduction method are reported in Table~\ref{tab:dr_time_memory}, which shows that PCA is fastest for reduction while VAE requires the most time for clustering, with all methods using similar memory during clustering phases. To ensure full reproducibility, all random seeds (NumPy, PyTorch, and scikit‐learn) were fixed at the start of each script, and the complete codebase alongside environment specifications is available on GitHub.

\begin{table}[ht]
\centering
\caption{Time and memory consumption statistics}
\label{tab:dr_time_memory}
\begin{tabular}{|p{1.1cm}|p{1cm}|p{1.8cm}|p{1cm}|p{1.8cm}|}
\hline
\textbf{Method} &
\multicolumn{2}{c|}{\textbf{Reduction}} &
\multicolumn{2}{c|}{\textbf{Clustering}} \\
\cline{2-5}
 & Time (s) & Memory (MB) &
   Time (s) & Memory (MB) \\
\hline 
PCA & 1.47 & 453 & 140 & 1499 \\
\hline
NMF & 66 & 459 & 277  & 1449 \\
\hline
AE & 60 & 607 & 282 & 1453 \\
\hline
VAE & 48 & 607 & 578 & 1437  \\
\hline
\end{tabular}
\vspace{-5mm}
\end{table}  
\section{Experiments\label{sec:expt}}
We compare six embeddings: (1) PCA, (2) NMF, (3) AE, (4) VAE, (5) RAW (clustering on the full (n$\times$ p) expression matrix without dimensionality reduction), and (6) hybrids (concatenated embeddings from PCA+NMF and VAE+NMF).
We conducted a full grid search over latent dimensionality $k$ and Leiden resolution $\rho$, evaluating all previously described metrics.
Section~\ref{sec:exp-sweeps} describes results of using Pareto optimal analysis to filter suboptimal hyperparameter configurations, and Section~\ref{sec:mer_ba} evaluates the before-versus-after impact of MER-based reassignment and presents a case study of a TMA sample.

\subsection{Pareto-front Analysis}
\label{sec:exp-sweeps}

\begin{figure*}[ht]
  \centering
  \begin{subfigure}[t]{0.4\textwidth}
    \centering
    \includegraphics[width=\linewidth]{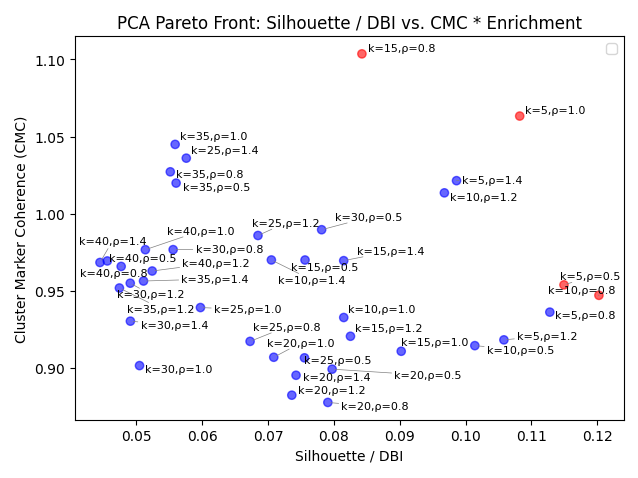}
    \caption{PCA Pareto front}
    \label{fig:pareto_pca}
  \end{subfigure}%
  \begin{subfigure}[t]{0.4\textwidth}
    \centering
    \includegraphics[width=\linewidth]{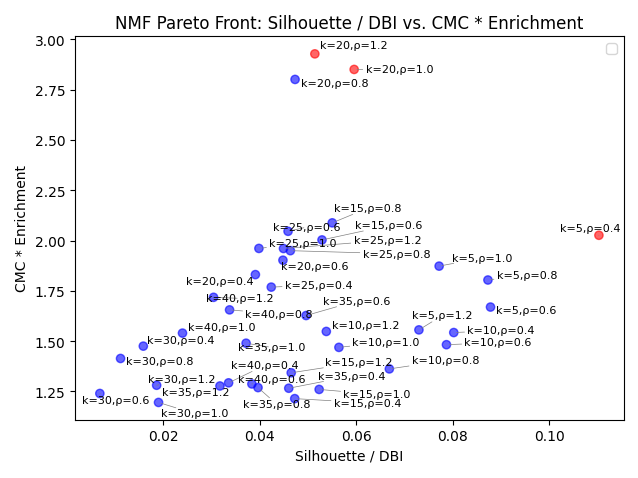}
    \caption{NMF Pareto front}
    \label{fig:pareto_nmf}
  \end{subfigure}

  \vspace{1em}

  \begin{subfigure}[t]{0.4\textwidth}
    \centering
    \includegraphics[width=\linewidth]{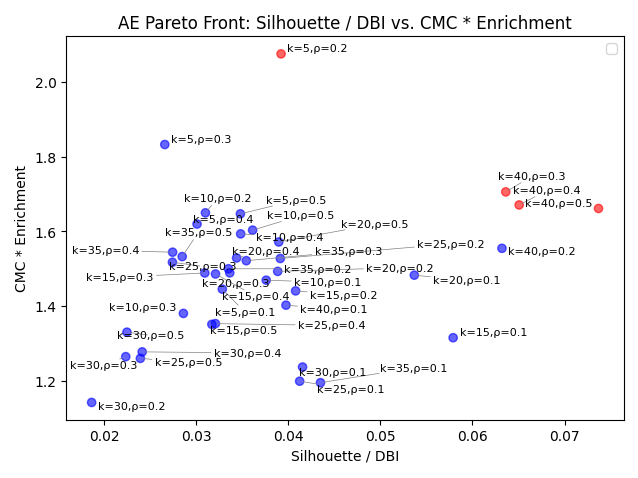}
    \caption{AE Pareto front}
    \label{fig:pareto_ae}
  \end{subfigure}%
  \begin{subfigure}[t]{0.4\textwidth}
    \centering
    \includegraphics[width=\linewidth]{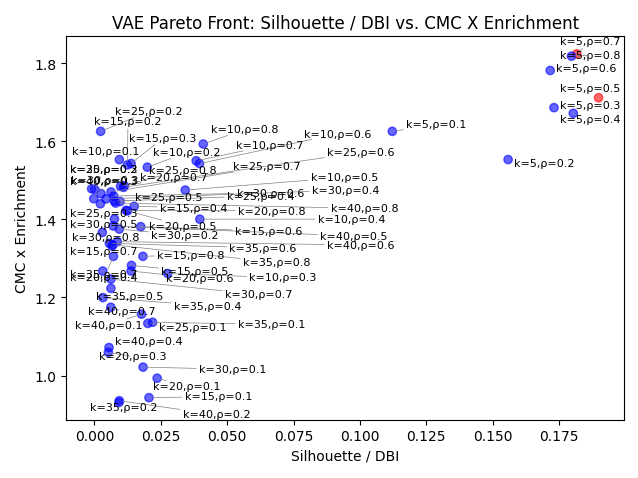}
    \caption{VAE Pareto front}
    \label{fig:pareto_vae}
  \end{subfigure}

  \caption{Pareto fronts (Silhouette/ DBI vs.\ CMC * Enrichment) for each dimensionality‐reduction method.  Red points denote Pareto‐optimal hyperparameter settings.}
  \label{fig:pareto_all}
\end{figure*}

Figure~\ref{fig:pareto_all} displays the pareto fronts observed using methods PCA, NMF, AE, and VAE by plotting the ratio of Silhouette to Davies–Boul-din index on the horizontal-axis against the product of CMC and enrichment score on the vertical-axis, with each point labeled by its latent dimension \(k\) and resolution \(\rho\).
In each panel, red points highlight configurations for which no other run outperforms both metrics simultaneously, revealing the trade‐off frontier between clustering compactness and biological coherence across methods.  

Table~\ref{tab:best_configs} reports, for each dimensionality‐reduction method, all hyperparameter settings on the Pareto curve (latent dimension $k$ and Leiden resolution $\rho$. We also list the corresponding Davies–Bouldin index (DBI), cluster‐marker coherence (CMC), mean‐squared error (MSE), explained variance, enrichment score, marker exclusion rate (MER), and marker-zero cells (MZC) for each best configuration. 

\begin{table*}[ht]
\small
\centering
\caption{Pareto-optimal hyperparameter settings per method and their performance metrics.}
\label{tab:best_configs}
\begin{tabular}{|l|p{0.7cm}|p{0.6cm}|p{1cm}|p{1.5cm}|p{1cm}|p{1cm}|p{1.8cm}|p{1.4cm}|p{1.2cm}|p{1cm}|p{1cm}|}
\hline
Method 
  & $k$
  & $\rho$
  & Clusters
  & Silhouette $\uparrow$
  & DBI $\downarrow$
  & CMC $\uparrow$
  & Recon. Error $\downarrow$
  & Expl. Var. $\uparrow$ 
  & Enrich. $\uparrow$
  & MER $\downarrow$
  & MZC $\downarrow$ \\
\hline
\hline
PCA  
  & 5 
  & 0.5  
  & 16
  & 0.168 
  & 1.463  
  & 0.783  
  & 2.577  
  & -0.289
  & 1.219  
  & 0.599
  & 52 \\
\hline
PCA  
  & 5 
  & 1.0  
  & 28
  & 0.152 
  & 1.408  
  & 0.795  
  & 2.577  
  & -0.289
  & 1.338  
  & 0.716
  & 1 \\
\hline
PCA  
  & 10 
  & 0.8  
  & 15
  & 0.188 
  & 1.559  
  & 0.819  
  & 2.472  
  & -0.237
  & 1.156  
  & 0.426 
  & 44 \\
\hline
PCA  
  & 15 
  & 0.8  
  & 22
  & 0.148 
  & 1.755  
  & 0.844  
  & 2.398  
  & -0.2
  & 1.308  
  & 0.309
  & 1 \\
\hline
\hline
NMF 
  & 5 
  & 0.4  
  & 19 
  & 0.150 
  & 1.357  
  & 0.772
  & 1.595
  & 0.202  
  & 2.626 
  & 0.661 
  & 139 \\
\hline
NMF 
  & 20 
  & 1.0  
  & 55 
  & 0.120 
  & 2.013  
  & 0.843
  & 1.401
  & 0.299  
  & 3.380 
  & 0.730
  & 26 \\
\hline
NMF 
  & 20 
  & 1.2  
  & 61 
  & 0.108 
  & 2.095  
  & 0.841
  & 1.401
  & 0.299  
  & 3.484 
  & 0.735
  & 18 \\
\hline
\hline
AE 
  & 5  
  & 0.2  
  & 36 
  & 0.063  
  & 1.615  
  & 0.729  
  & 1.602 
  & 0.198
  & 2.845
  & 0.577
  & 2 \\
\hline
AE 
  & 40  
  & 0.3  
  & 40 
  & 0.104  
  & 1.631  
  & 0.737  
  & 1.600 
  & 0.199
  & 2.313
  & 0.685
  & 0 \\
\hline
AE 
  & 40  
  & 0.4  
  & 48 
  & 0.104  
  & 1.600  
  & 0.752  
  & 1.600 
  & 0.199
  & 2.220
  & 0.567
  & 0 \\
\hline
AE 
  & 40  
  & 0.5  
  & 51 
  & 0.115  
  & 1.558  
  & 0.750  
  & 1.600 
  & 0.199
  & 2.214
  & 0.580
  & 0 \\
\hline
\hline
VAE 
  & 5  
  & 0.5  
  & 49
  & 0.237  
  & 1.250 
  & 0.764 
  & 1.457  
  & 0.271
  & 2.242  
  & 0.478 
  & 0\\
\hline
VAE 
  & 5  
  & 0.7  
  & 56
  & 0.228  
  & 1.255 
  & 0.763 
  & 1.457  
  & 0.271
  & 2.390  
  & 0.652 
  & 0 \\
\hline
\hline
RAW 
  & -  
  & 0.2  
  & 14
  & 0.044  
  & 4.056 
  & 0.857 
  & 0.000  
  & 1.000
  & 2.907  
  & 0.343
  & 341 \\
\hline
RAW 
  & -  
  & 0.4  
  & 17
  & 0.047  
  & 3.933 
  & 0.864 
  & 0.000  
  & 1.000
  & 2.355  
  & 0.391
  & 52 \\
\hline
\hline
PCA+NMF 
  & 5, 10  
  & 0.2  
  & 8
  & 0.261  
  & 1.120 
  & 0.822 
  & 1.759  
  & 0.120
  & 1.647  
  & 0.160
  & 290 \\
\hline
PCA+NMF 
  & 10, 10  
  & 0.3  
  & 11
  & 0.222  
  & 1.364 
  & 0.846 
  & 1.170  
  & 0.144
  & 1.410  
  & 0.249 
  & 191 \\
\hline
\hline
VAE+NMF 
  & 5, 5  
  & 0.4  
  & 34
  & 0.184  
  & 1.621 
  & 0.776 
  & 1.470  
  & 0.264
  & 2.140  
  & 0.405
  & 0 \\
\hline
VAE+NMF 
  & 5, 5  
  & 0.5 
  & 38
  & 0.179  
  & 1.682 
  & 0.779 
  & 1.470  
  & 0.264
  & 2.277  
  & 0.389
  & 0 \\
\hline
VAE+NMF 
  & 5, 5  
  & 0.6  
  & 39
  & 0.141  
  & 1.642 
  & 0.777 
  & 1.470  
  & 0.264
  & 2.214  
  & 0.513
  & 0\\
\hline
VAE+NMF 
  & 10, 15  
  & 0.1  
  & 11
  & 0.172  
  & 1.668 
  & 0.814 
  & 1.367  
  & 0.316
  & 3.150  
  & 0.286
  & 375\\
\hline
\hline
\end{tabular}
\end{table*}

Below are eight principal takeaways distilled from the Pareto‐optimal plots~\ref{fig:pareto_all} and the summary in Table~\ref{tab:best_configs}.

\textbf{1. VAE strikes the best overall balance}: Among single methods, the VAE often lives on the extreme upper‐right of the Silhouette/DBI vs CMC×Enrichment plane (e.g. $k=5$, $\rho=0.5$, Silhouette $=0.237$, Enrichment $=2.242$), achieving both tight clusters and strong marker enrichment. It also shows moderate MER ($\approx$ $0.48$–$0.65$), indicating the fraction of cells that would need to be reassigned to clusters whose marker genes they express more strongly than those of their assigned cluster.

\textbf{2. PCA still delivers very cohesive clusters:} By ``cohesive clusters" we mean groups of cells whose profiles lie close together in the reduced-dimensional embedding, indicating tight, well-defined groupings. PCA’s top Pareto points (e.g.\ $k=10$, $\rho=0.8$, Silhouette $=0.188$, CMC $=0.819$) yield the highest Silhouette scores, reflecting the clearest separation in $Z$. However, PCA’s biological fidelity (CMC, enrichment) and MER ($\approx$ $0.31$–$0.72$, typical $\approx$ $0.6$-$0.7$) lag behind VAE and NMF.

\textbf{3. NMF excels at biological signal recovery:} NMF’s embeddings group cells closely aligning with canonical marker‐gene signatures. NMF's top Pareto settings (e.g.\ $k=20$, $\rho=1.2$, CMC $=0.841$, Enrichment $=3.48$) yield the highest marker‐set enrichment, surpassing all other single methods. Therefore, NMF produces clusters that not only perform well on numerical metrics but also correspond cleanly to biologically defined cell types, making it the method of choice when faithful recovery of known marker‐based identities is paramount. Its cluster tightness is moderate (Silhouette $\approx0.108$–$0.150$) and its MER ($\approx0.66$–$0.74$) is on par with PCA, underscoring that NMF trades some pure geometric separation for superior biological coherence.

\textbf{4. AE offer a middle ground:} AEs yield intermediate Silhouette ($\approx$ $0.063$-$0.115$) and CMC ($\approx$ $0.729$-$0.752$) scores with enrichment of ($\approx$ $2.214$-$2.845$) and MER $\approx$ $0.567$-$0.685$. They are particularly advantageous when one needs to capture nonlinear structure or denoise noisy expression profiles. Both AE and VAE models require careful hyperparameter tuning, so automating their training for non-expert users is an important avenue for future work. 

\textbf{5. Hybrid embeddings (PCA + NMF or VAE + NMF) can improve trade-offs:} For example, PCA + NMF (PCA=10, NMF=10, at resolution $\rho$=0.3) boosts Silhouette to $0.222$ and CMC to $0.846$ with MER $\approx$ $0.26$, which is better than PCA alone. VAE + NMF hybrids capture the high enrichment of NMF and the cohesion of VAE.

\textbf{6. Marker Exclusion Rate (MER) reveals mis-assignment risk:} MER correlates inversely with CMC and enrichment, and methods with high marker coherence have MER in the lower-0.4's, indicating higher fraction of cells expressing their cluster’s markers. On the other hand, higher MER ($\approx$ $0.6$–$0.8$) suggests losing biologically meaningful assignments.

\textit{Almost all cells match at least one marker set:} Across all methods, fewer than 0.2 \% of cells express \emph{none} of the cluster’s markers (MZC < 0.2\%), confirming that nearly all cells fall into at least one cluster representing a biologically coherent group.
    
\textbf{7. Dimension reduction is essential:}  Clustering the full \(n\times p\) matrix yields Silhouette \(\approx0.04\), DBI \(\approx4.0\), CMC \(\approx0.85\), zero reconstruction error, 100 \% explained variance, and MER \(\approx0.34\).  However, the clusters are loose and biologically ambiguous as indicated by these scores. For example, the low Silhouette score means that on average a cell is almost as close to its nearest ``other" cluster as it is to its own, the high DBI indicates that within‐cluster scatter is nearly as large as (or larger than) between‐cluster separation. Then although CMC is high, MER shows that over a third of cells don’t express the markers of their assigned cluster, so many clusters lack clear biological identity. All of this underscores the fact that dimensionality reduction is a key ingredient for obtaining tight, and biologically coherent cell groupings.
    
\textbf{8. No one-size-fits-all, choose by one's priority:} If the goal is purely geometric cluster cohesion, PCA’s top Pareto points give the tightest, most well‐separated clusters in reduced space, though at the cost of higher marker‐misassignment (MER). Because PCA maximizes global variance along orthogonal axes, it naturally concentrates cells into compact, well-separated clusters in reduced space, even when those variance directions don’t align with individual marker genes. So subtle, marker-specific distinctions get lost.
    
    If biological fidelity (marker recovery) is paramount, NMF is the preferred choice, with VAE offering a balanced trade-off between reconstruction accuracy and clustering performance. Alternatively, hybrid approaches (e.g., PCA + NMF concatenation) can capture complementary strengths, preserving PCA's geometric cohesion while enhancing marker recovery through NMF's parts-based decomposition.

\subsection{MER-guided Post Processing\label{sec:mer_ba}}
We selected six Pareto-optimal representative configurations (from Table~\ref{tab:best_configs}) to study before-versus-after results of MER-guided post-processing. Table~\ref{tab:rep_configs} shows that MER-guided post-processing successfully drives all MER values to zero while improving CMC scores across methods. Although silhouette scores decrease compared to pre-processing results, this metric becomes less relevant since it evaluates the original embedding-based cluster assignments that are intentionally modified during post-processing to prioritize biological marker coherence over geometric proximity.
\begin{table}[ht]
\small
\centering
\caption{Representative Pareto‐optimal configurations after MER-guided cell reassignment.}
\label{tab:rep_configs}
\begin{tabular}{|l|c|c|c|c|c|c|c|}
\toprule
Method & $k$ & $\rho$ & clusters & Sil & CMC & Enrich & MER \\
\midrule
PCA          & 10 & 0.8 & 15 & 0.096 & 0.866 & 1.156 & 0 \\
NMF          & 20 & 1.2 & 61 & -0.133 & 0.901 & 3.484 & 0 \\
VAE          &  5 & 0.7 & 56 & -0.063 & 0.862 & 2.390 & 0 \\
AE           & 5 & 0.2 & 30 & -0.119 & 0.819 & 2.845 & 0 \\
PCA+NMF\,*   & 10,10 & 0.3 & 11 & 0.145
& 0.863 & 1.410 & 0 \\
VAE+NMF\,*   & 5,5 & 0.5 & 38 & 0.004 & 0.848 & 2.277 & 0 \\
\bottomrule
\end{tabular}
\\[1ex]
\footnotesize
\,*Hybrid embeddings formed by concatenating the two latent representations.
\end{table}

\eatme{
\begin{figure}[ht]
  \centering
  \includegraphics[width=0.4\textwidth]{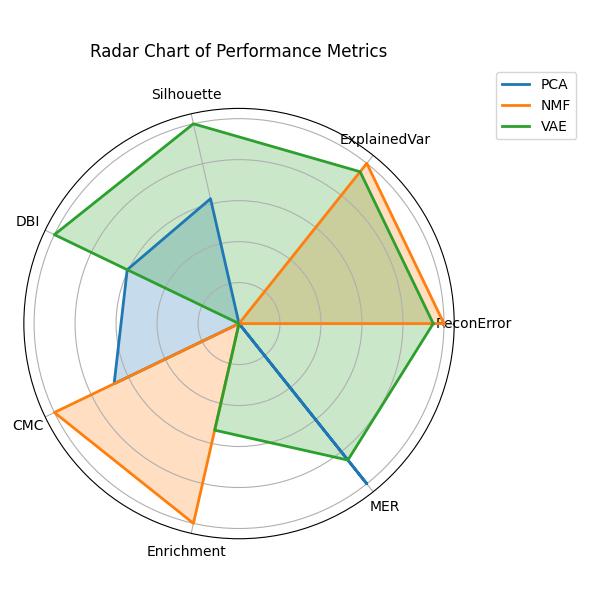}
  \caption{Radar chart comparing normalized performance metrics across dimensionality reduction methods (PCA, NMF, and VAE).  Metrics include reconstruction error (inverted), explained variance, silhouette, DBI (inverted), CMC, enrichment score, and MER (inverted).}
  \label{fig:radar_performance}
\end{figure}
}

\begin{figure}[t]
  \centering
  \includegraphics[width=\columnwidth]{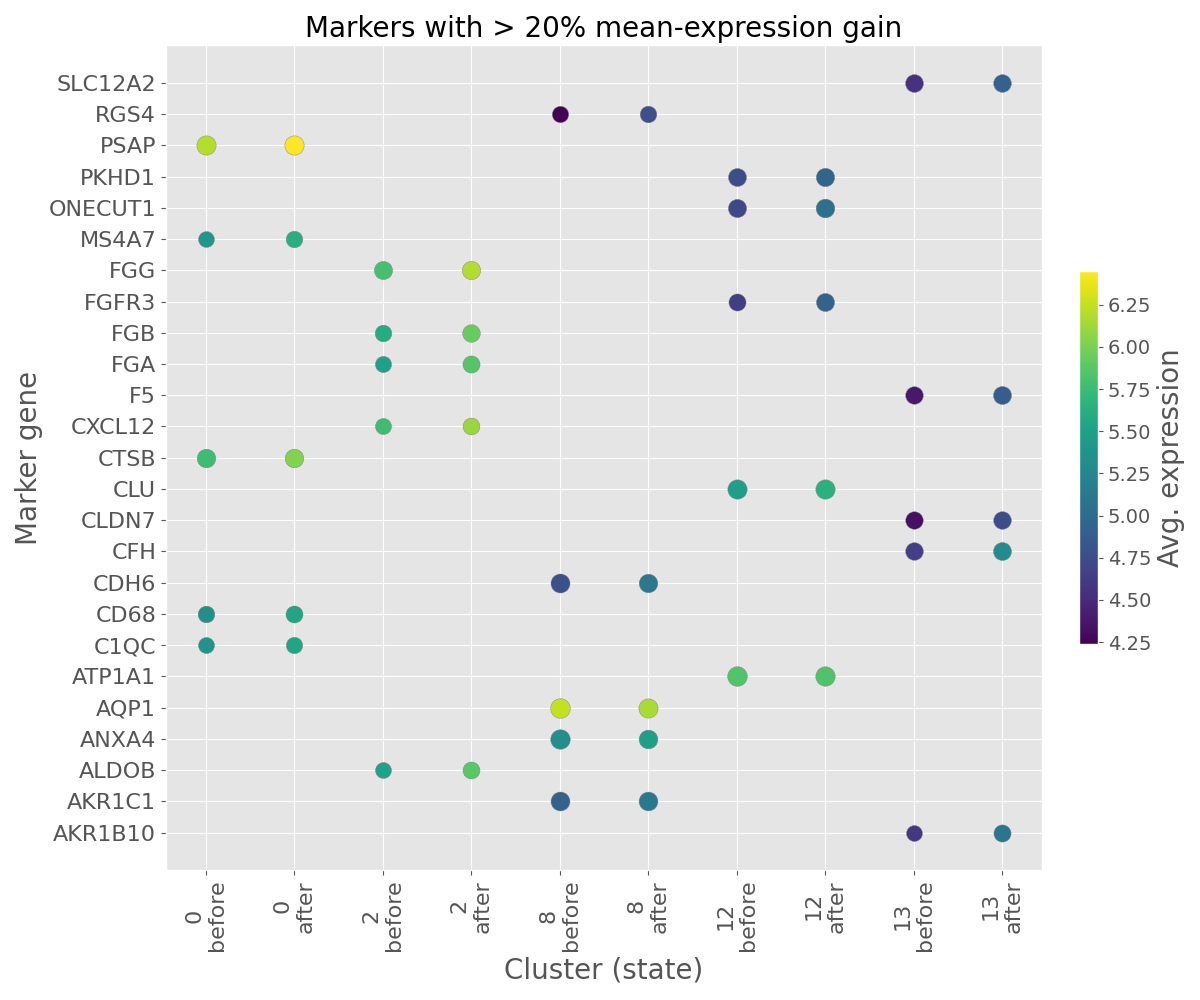}
  \caption{Markers with \(\boldsymbol{>20\%}\) mean‐expression gain after MER‐guided reassignment of the PCA configuration in Table~\ref{tab:rep_configs}.
    For each cluster we show before-versus-after states side by side (x–axis), and for each marker gene the point size encodes the fraction of positive cells while the color encodes average expression among positives.  Only those genes whose mean expression increased by more than 20 \% are shown.
  }
  \label{fig:marker_dotplot_subset}
  \vspace{-5mm}
\end{figure}

\begin{figure*}[htbp]
  \centering

  \begin{subfigure}[b]{0.24\textwidth}
    \includegraphics[clip,trim=90 0 90 10,width=\linewidth]{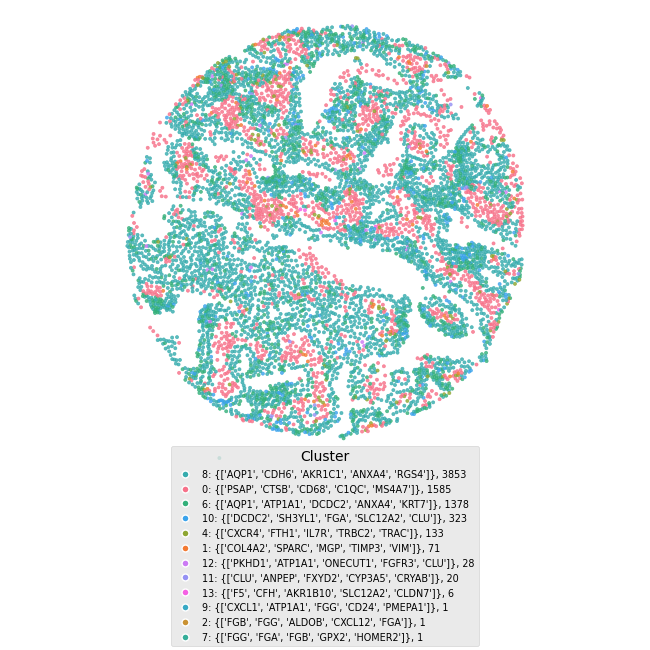}
    \caption{PCA Before}
    \label{fig:pca_before}
  \end{subfigure}
  \begin{subfigure}[b]{0.24\textwidth}
    \includegraphics[clip,trim=100 0 100 0,width=\linewidth]{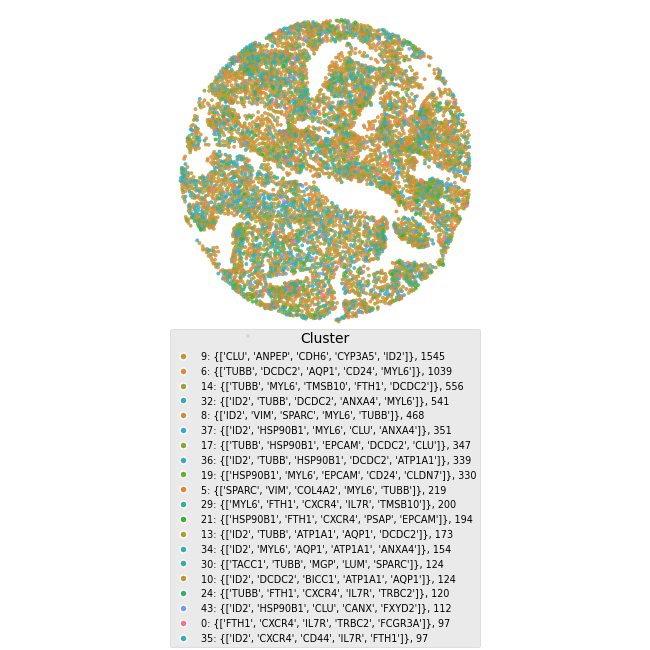}
    \caption{NMF Before}
    \label{fig:nmf_before}
  \end{subfigure}
    \begin{subfigure}[b]{0.24\textwidth}
    \includegraphics[clip,trim=100 0 100 0,width=\linewidth]{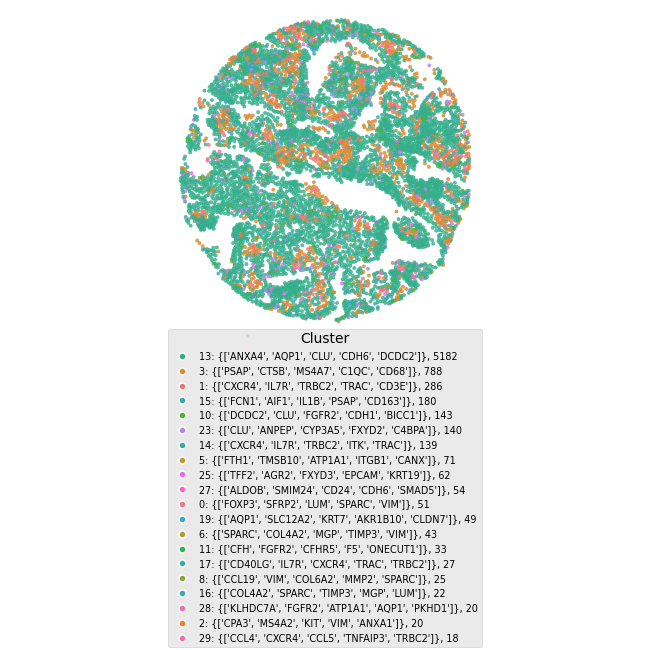}
    \caption{AE Before}
    \label{fig:ae_before}
  \end{subfigure}
    \begin{subfigure}[b]{0.24\textwidth}
    \includegraphics[clip,trim=100 0 100 0, width=\linewidth]{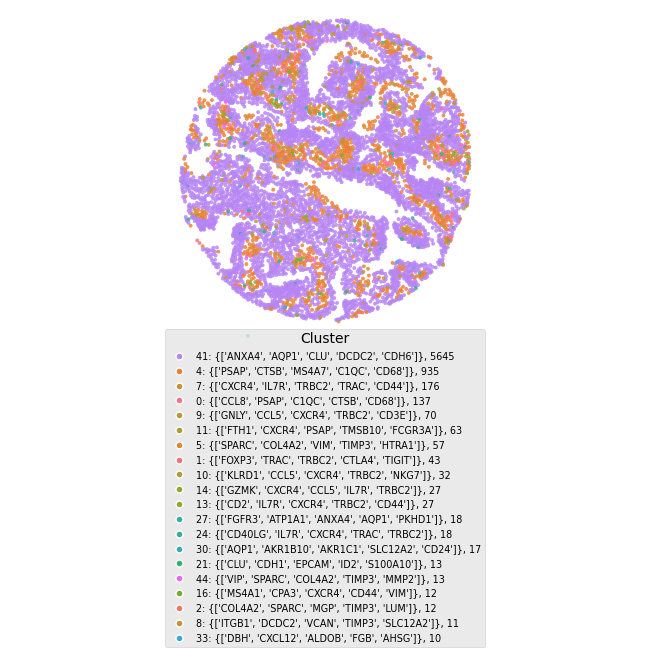}
    \caption{VAE Before}
    \label{fig:vae_before}
  \end{subfigure}


    \begin{subfigure}[b]{0.24\textwidth}
    \includegraphics[clip,trim=90 0 90 10,width=\linewidth]{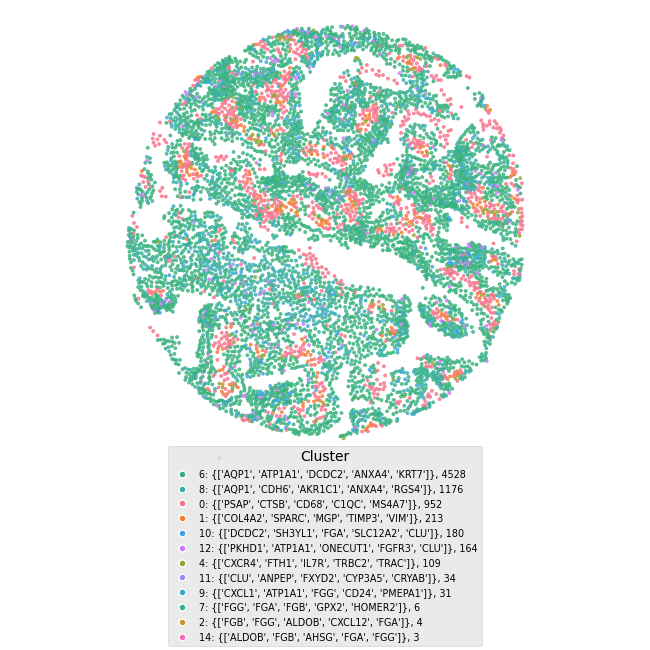}
    \caption{PCA After}
    \label{fig:pca_after}
  \end{subfigure}\hfill
  \begin{subfigure}[b]{0.24\textwidth}
    \includegraphics[clip,trim=100 0 100 0,width=\linewidth]{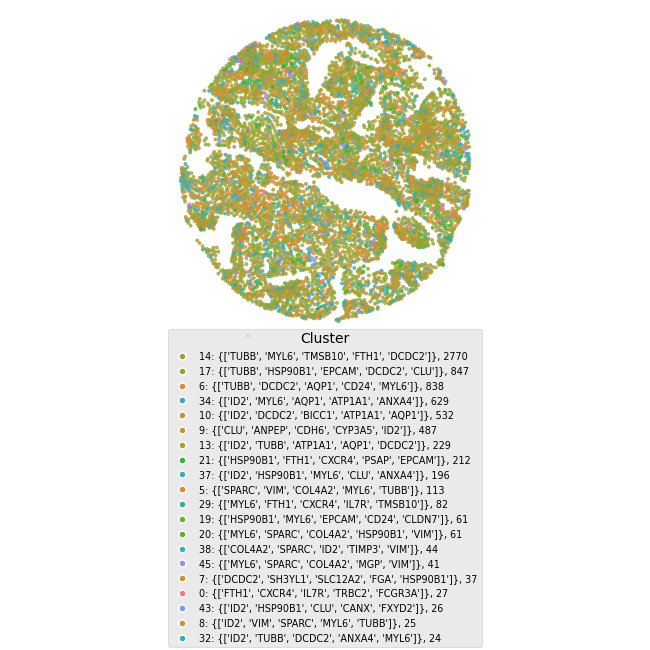}
    \caption{NMF After}
    \label{fig:nmf_after}
  \end{subfigure}
  \begin{subfigure}[b]{0.24\textwidth}
    \includegraphics[clip,trim=100 0 100 0,width=\linewidth]{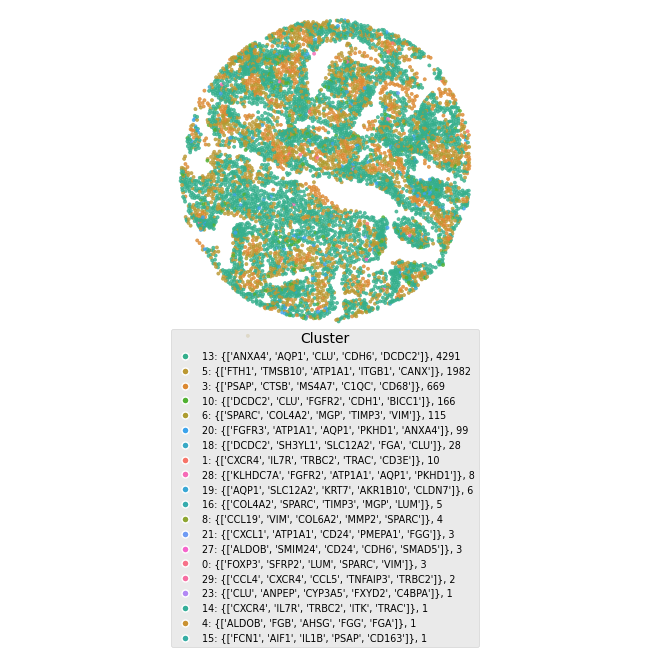}
    \caption{AE After}
    \label{fig:ae_after}
  \end{subfigure}\hfill
  \begin{subfigure}[b]{0.24\textwidth}
    \includegraphics[clip,trim=100 0 100 0, width=\linewidth]{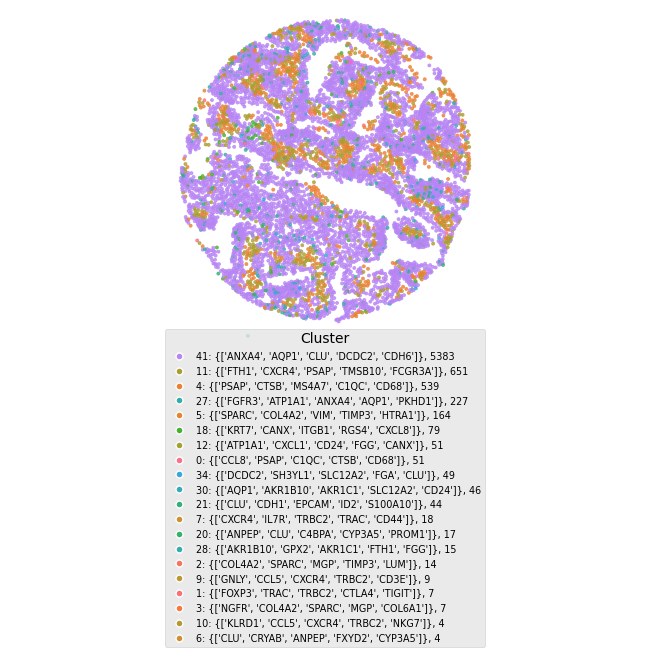}
    \caption{VAE After}
    \label{fig:vae_after}
  \end{subfigure}

  \caption{Cell-type atlassing results for PCA, NMF, AE, VAE. Each column shows one method; the top row displays atlassing *before* cluster-label reassignment, and the bottom row shows the
same embeddings *after* reassignment based on MER-driven post processing. The legend gives the cluster number, the top-5 marker genes, and the number of cells in a cluster.}
  \label{fig:express_grid}
\end{figure*}

Figure~\ref{fig:marker_dotplot_subset} demonstrates the effectiveness of MER-guided post-processing by showing markers that gained $>20$ \% mean expression after reassignment in the PCA configuration. The before-versus-after comparison reveals that post-processing successfully enhanced marker expression within their appropriate clusters, with both the fraction of positive cells (point size) and average expression levels (color intensity) improving across multiple marker genes and cluster pairs.

Figure~\ref{fig:express_grid} shows a case study across four dimensionality-reduction pipelines applied to the same cholangiocarcinoma TMA sample. We consistently recovered four dominant gene modules representing the primary biological cell populations in the specimen. A large cholangiocyte signature, characterized by AQP1, ANXA4, CDH6, DCDC2, CLU, was present in every embedding, but its granularity varied: PCA and NMF split the epithelium into several sub-clusters, whereas AE and especially VAE compressed it into a single, cohesive manifold. All methods also identified a monocytic/macrophage module (PSAP, CTSB, MS4A7, C1QC, CD68), although the non-linear models isolated an additional FCN1/AIF1-high inflammatory subset that remained conflated in PCA. A stromal/mesenchymal module (SPARC, VIM, COL4A2, MGP, TIMP3) was reproducible across methods; VAE further distinguished a small MMP-rich fibroblast population, suggesting that the stochastic latent space is sensitive to rare extracellular-matrix states. Finally, a lymphoid cluster bearing CXCR4, IL7R, TRBC2, TRAC, CD3E emerged in every representation but was most sharply delineated in NMF, reflecting that parts-based factors accentuate discrete immune modules. Taken together, the concordance of core marker sets confirms the biological robustness of the atlas, while the method-specific differences highlight how algorithmic choices modulate the resolution at which epithelial, stromal and immune heterogeneity is revealed.

Analyzing the images in the bottom row of Figure~\ref{fig:express_grid}, we find that MER-guided reassignment collapsed method-specific artefacts and revealed clearer functional gene signatures across methods. For example, the autoencoder's refined clustering now shows distinct cytoskeletal (TUBB, MYL6), metabolic (HSP90B1, SPARC), and signaling (CXCR4, TMSB10) modules that were previously obscured by over-clustering. VAE's post-processing maintained its core epithelial signature (ANXA4, AQPQ1, CLU, DCDC2) while better defining specialized subpopulations with markers like FGFR3 and KRT7. The refinement appears to have resolved cases where cells were misassigned based on embedding proximity rather than actual gene expression patterns.

Importantly, MER-guided post-processing improved clustering quality regardless of the underlying dimensionality reduction method. This suggests the approach addresses a fundamental limitation of embedding-based clustering—that cells can be grouped by latent space proximity despite expressing markers more characteristic of other clusters. The consistent improvements across PCA, NMF, autoencoder, and VAE demonstrate that marker-guided refinement provides a robust, method-agnostic solution for achieving more biologically coherent cell type assignments.

\eatme{
In fact, now four coherent gene programs now dominate the map irrespective of the dimensionality-reduction method. These are listed below:

\paragraph*{Cholangiocyte core programme: ANXA4, AQP1, CLU, CDH6, DCDC2} 
MER merges the diffuse epithelial fragments that PCA and NMF had split and that AE/VAE had partly blended. Its marker list is now more consistent across dimensionality reduction techniques, highlighting a stable ductal identity.

\paragraph*{Resident macrophage / monocyte module: PSAP, CTSB, MS4A7, C1QC, CD68}
In the raw partitions this module was diluted by stress genes; MER expels the contaminants and yields compact clusters (#0 PCA, #3 AE, #4 VAE, #21 NMF) whose top-5 markers fully agree. The macrophage set therefore survives both linear and non-linear dimensionality reduction once noisy genes are down-weighted.

\paragraph*{Stromal–mesenchymal axis: SPARC,  COL4A2, VIM, MGP, TIMP3} 
A previously fragmented fibroblast population coalesces into a single territory (#1 PCA, #6 NMF, #6 AE, #5 VAE). MER suppresses small VIM-low offshoots and aligns the stromal gene loadings, giving the same five ECM markers maximal enrichment in all four methods.

\paragraph*{Adaptive/innate lymphoid compartment: CXCR4, IL7R, TRBC2, TRAC, CD3E} with FTH1/TMSB10 support genes.
T cells, which PCA and NMF had partially mixed with macrophages and AE/VAE had under-segmented, now form crisp islands (#4 PCA, #17 NMF, #11 VAE, #5 AE). The shared marker set illustrates MER’s capacity to retrieve low-abundance lineages without over-splitting them.
}

\subsection{Discussion\label{sec:discussion}}
\label{sec:exp-discussion}
We compared four dimensionality-reduction techniques (PCA, NMF, AE, VAE), raw expression, and two hybrids (PCA + NMF, VAE + NMF) on spatial transcriptomics data using standard clustering metrics (Silhouette score, Davies–Bouldin index) and two novel biological measures: Cluster Marker Coherence (CMC) and Marker Exclusion Rate (MER).
PCA produced the tightest clusters (highest Silhouette) but failed to recover canonical marker signatures (lower CMC/enrichment, higher MER). NMF balanced moderate cohesion with strong marker enrichment, mapping cleanly onto known cell types despite lower silhouette scores. Autoencoders (AE, VAE) captured nonlinear structure and denoised data with respectable performance but required careful tuning.
No embedding was perfect: 30–80\% of cells were mis-assigned to clusters whose markers they don't express best. We introduced MER-guided post-processing to reassign cells based on summed marker expression, leveraging original gene data to correct clustering labels. After MER-reassignment, dimensionality-reduction choice mainly impacts marker-gene signature survival, making biological fidelity the primary differentiator.

MER reassignment can be enhanced through three complementary refinements. First, \textit{adaptive merging thresholds} that scale with cluster size and marker entropy would prevent niche populations (e.g., FCN1/AIF1-high inflammatory macrophages) from being absorbed into resident macrophages, while still collapsing spurious splits within major epithelial compartments. Second, \textit{marker-weight overrides} would allow users to whitelist protected genes or blacklist ubiquitous housekeeping genes, thereby safeguarding biologically meaningful clusters and eliminating technical artefacts. Third, \textit{integrating MER with spatial information} could veto merges that conflate biologically incompatible cells. Together, these refinements would transform MER from a generic clean-up step into a tunable, context-aware post-processor that preserves rare states while delivering concise, interpretable atlases.

\section{Conclusion\label{sec:conclude}}
We systematically evaluated non-spatially aware dimensionality-reduction methods (PCA, NMF, AE, VAE) on a spatial transcriptomics benchmark. Our key contributions are:
\begin{compactenum}
\item Cluster Marker Coherence (CMC): a metric quantifying the average fraction of cells within each cluster that express that cluster's canonical marker genes.
\item Marker Exclusion Rate (MER): a complementary metric measuring the fraction of cells whose marker-gene profile would better suit a different cluster.
\item MER-guided reassignment: a post-processing step that corrects cluster labels by reassigning cells to the cluster whose markers they express most strongly. This procedure substantially improves both CMC and MER, and boosts average marker-gene expression, enhancing downstream analyses such as niche identification and cell-cell interaction mapping.
\end{compactenum}
While no single method dominates, our results confirm that PCA excels in geometric cohesion (high Silhouette, low DBI), whereas NMF produces the most biologically coherent clusters (high CMC, enrichment). Autoencoders (AE, VAE) offer a flexible, nonlinear alternative but require careful tuning. Importantly, MER-guided reassignment levels the playing field: after correction, the main differentiator among methods is the fidelity of their recovered marker signatures rather than clustering tightness.
In future work, we will extend this framework to spatially aware dimensionality reduction techniques, incorporating local tissue architecture into both embedding and evaluation, and assess their impact on spatial-omics workflows across multiple platforms.

\vspace{-2mm}
\section*{Acknowledgment}
The work was, in part, supported by STRIDE funding to METI from MD Anderson Cancer Center. We thank the cholangiocarcinoma clinical and pathology group at MDACC for TMA. 
\vspace{-2mm}

\eatme{
\section*{Per‐Gene Reconstruction Error Analysis}
\begin{table*}[ht]
\centering
\caption{Top 10 Genes with Highest Reconstruction Errors by Method}
\label{tab:top_error_genes}
\begin{threeparttable}
\begin{tabular}{|l |p{0.65\textwidth}| c|}
\hline
Method
  & Top 10 Genes (MSE) 
  & \# Marker Overlap \\
\hline
PCA 
  & CLU (10.066), AQP1 (9.405), VIM       (8.247), 
DCDC2     (7.868),
KRT7      (6.505),
ANPEP     (5.395),
BICC1     (5.323),
PARK7     (5.054),
SPARC     (4.739),
TACC1     (4.573)  
  & 10 \\
\hline
NMF 
  & PARK7      (3.978),
CCL2       (2.703),
HTRA1      (2.702),
TNFAIP3    (2.678),
CFH        (2.620),
LY6E       (2.587),
GPX2       (2.572),
AKR1B10    (2.571),
ANXA13     (2.541),
SLPI       (2.443)
  & 5 \\
\hline
AE  
  & AQP1     (7.726),
CLU      (7.490),
FGG      (7.185),
DCDC2    (6.658),
KRT7     (6.298),
ANPEP    (6.214),
VIM      (6.095),
CD44     (5.817),
BICC1    (5.746),
PARK7    (5.569)
  & 9 \\
\hline
VAE 
  & VIM       (5.115),
PARK7     (5.082),
CD44      (4.936),
TIMP1     (4.738),
TACC1     (4.676),
AQP1      (4.510),
GRN       (4.125),
ANPEP     (4.113),
COL4A2    (4.033),
KRT7      (3.900)
  & 7 \\
\hline
\end{tabular}

\begin{tablenotes}[flushleft]
\footnotesize
\item[\#] AKR1B10: aldo-keto reductase family 1 member B10 (detoxifying enzyme for reactive aldehydes); ANPEP: alanyl (membrane) aminopeptidase, also known as CD13 (epithelial and stromal cell surface marker); ANXA13: annexin A13 (calcium-dependent phospholipid-binding protein involved in epithelial membrane organization and vesicle trafficking); AQP1: aquaporin-1 (endothelial marker); BICC1: bicaudal C homolog 1 (RNA-binding protein influencing cell polarity and ciliogenesis); CCL2: chemokine (C-C motif) ligand 2, also known as MCP-1 (monocyte chemoattractant protein-1) (macrophage/monocyte recruitment chemokine); CD44: cluster of differentiation 44 (cell–surface glycoprotein involved in cell–matrix adhesion and migration); CFH: complement factor H (regulator of the alternative complement pathway); CLU: clusterin (chaperone function); COL4A2: collagen type IV alpha 2 chain (basement membrane structural protein involved in extracellular matrix integrity and cell adhesion); DCDC2: doublecortin domain containing 2 (microtubule-stabilizing/ciliary function); FGG: fibrinogen gamma chain (coagulation protein and acute-phase reactant involved in blood clot formation); GRN: granulin precursor (progranulin) (growth factor involved in cell proliferation, wound repair, and inflammatory modulation); GPX2: glutathione peroxidase 2 (antioxidant enzyme involved in mucosal protection and oxidative stress response); HTRA1: HtrA serine peptidase 1 (serine protease involved in extracellular matrix remodeling and TGF-$\beta$ signaling); KRT7: keratin-7 (biliary epithelial intermediate‐filament marker); LY6E: lymphocyte antigen 6E (GPI-anchored immune modulation marker); PARK7: Parkinsonism associated deglycase DJ-1 (oxidative stress response chaperone); SLPI: secretory leukocyte protease inhibitor (serine protease inhibitor that modulates inflammatory responses and protects mucosal surfaces);  SPARC: osteonectin (matrix remodeling); TACC1: transforming acidic coiled-coil containing protein 1 (mitotic spindle regulator with roles in ECM interactions); TIMP1: tissue inhibitor of metalloproteinases 1 (regulator of extracellular matrix remodeling); TNFAIP3: TNF alpha induced protein 3, also known as A20 (ubiquitin-editing enzyme that dampens NF-$\kappa$B signaling); VIM: vimentin (mesenchymal marker);.
\end{tablenotes}
\end{threeparttable}
\end{table*}
Table~\ref{tab:top_error_genes} catalogs the ten genes with the highest reconstruction errors for each method and reports how many known cholangiocarcinoma markers appear in each top-10 list. Strikingly, even PCA’s worst reconstructions include all ten canonical markers, indicating that the genes we most need to preserve—rather than noise—are systematically distorted when compressed into a small number of components. The autoencoder similarly flags nine markers among its highest-error genes, and the VAE and standard NMF recover only seven and five markers, respectively, underscoring a shared inability across methods to faithfully represent critical lineage and functional genes in low-dimensional space.

A core set of genes—AQP1, CLU, VIM, DCDC2, KRT7, ANPEP, PARK7, SPARC, and TACC1—appears repeatedly among the highest-error lists, highlighting the challenge of embedding features with complex, multimodal regulation and heterogeneous spatial patterns. Standard NMF in particular struggles with immune and inflammatory mediators (e.g.\ CCL2, TNFAIP3, CFH, GPX2, SLPI), while the VAE’s top errors include matrix-remodeling genes (CD44, TIMP1), reflecting each method’s bias toward reconstructing overall variance at the expense of biologically essential signals.

These results reveal a fundamental trade-off: aggressive dimensionality reduction inevitably sacrifices fidelity for the very genes that define cell identity, stress response, extracellular matrix remodeling, and immune regulation. Addressing this gap 
will be essential to ensure that low-dimensional representations retain the nuanced biology at the heart of cholangiocarcinoma heterogeneity.
}

\bibliographystyle{ACM-Reference-Format}  
\bibliography{sample-base} 


\begin{thebibliography}{23}


\ifx \showCODEN    \undefined \def \showCODEN     #1{\unskip}     \fi
\ifx \showISBNx    \undefined \def \showISBNx     #1{\unskip}     \fi
\ifx \showISBNxiii \undefined \def \showISBNxiii  #1{\unskip}     \fi
\ifx \showISSN     \undefined \def \showISSN      #1{\unskip}     \fi
\ifx \showLCCN     \undefined \def \showLCCN      #1{\unskip}     \fi
\ifx \shownote     \undefined \def \shownote      #1{#1}          \fi
\ifx \showarticletitle \undefined \def \showarticletitle #1{#1}   \fi
\ifx \showURL      \undefined \def \showURL       {\relax}        \fi
\providecommand\bibfield[2]{#2}
\providecommand\bibinfo[2]{#2}
\providecommand\natexlab[1]{#1}
\providecommand\showeprint[2][]{arXiv:#2}

\bibitem[Bishop(2006)]%
        {Bishop2006}
\bibfield{author}{\bibinfo{person}{Christopher~M. Bishop}.} \bibinfo{year}{2006}\natexlab{}.
\newblock \bibinfo{booktitle}{\emph{Pattern Recognition and Machine Learning (Information Science and Statistics)}}.
\newblock \bibinfo{publisher}{Springer-Verlag}, \bibinfo{address}{Berlin, Heidelberg}.
\newblock
\showISBNx{0387310738}


\bibitem[Brunet et~al\mbox{.}(2004)]%
        {brunet2004nmf}
\bibfield{author}{\bibinfo{person}{Jean-Philippe Brunet}, \bibinfo{person}{Pablo Tamayo}, \bibinfo{person}{Todd~R Golub}, {and} \bibinfo{person}{Jill~P Mesirov}.} \bibinfo{year}{2004}\natexlab{}.
\newblock \showarticletitle{Metagenes and molecular pattern discovery using matrix factorization}.
\newblock \bibinfo{journal}{\emph{Proceedings of the National Academy of Sciences}} \bibinfo{volume}{101}, \bibinfo{number}{12} (\bibinfo{year}{2004}), \bibinfo{pages}{4164--4169}.
\newblock


\bibitem[Du{\`o} et~al\mbox{.}(2020)]%
        {duo2020systematic}
\bibfield{author}{\bibinfo{person}{Angelo Du{\`o}}, \bibinfo{person}{Mark~D Robinson}, {and} \bibinfo{person}{Charlotte Soneson}.} \bibinfo{year}{2020}\natexlab{}.
\newblock \showarticletitle{A systematic performance evaluation of clustering methods for single-cell RNA-seq data}.
\newblock \bibinfo{journal}{\emph{F1000Research}}  \bibinfo{volume}{7} (\bibinfo{year}{2020}), \bibinfo{pages}{1141}.
\newblock


\bibitem[Eraslan et~al\mbox{.}(2019)]%
        {eraslan2019autoencoders}
\bibfield{author}{\bibinfo{person}{G{\"o}kcen Eraslan}, \bibinfo{person}{Lukas~M Simon}, \bibinfo{person}{Merve Mircea}, \bibinfo{person}{Nadine~S Mueller}, {and} \bibinfo{person}{Fabian~J Theis}.} \bibinfo{year}{2019}\natexlab{}.
\newblock \showarticletitle{Single-cell RNA-seq denoising using a deep count autoencoder}.
\newblock \bibinfo{journal}{\emph{Nature communications}} \bibinfo{volume}{10}, \bibinfo{number}{1} (\bibinfo{year}{2019}), \bibinfo{pages}{1--14}.
\newblock


\bibitem[Glorot and Bengio(2010)]%
        {pmlr-v9-glorot10a}
\bibfield{author}{\bibinfo{person}{Xavier Glorot} {and} \bibinfo{person}{Yoshua Bengio}.} \bibinfo{year}{2010}\natexlab{}.
\newblock \showarticletitle{Understanding the difficulty of training deep feedforward neural networks}. In \bibinfo{booktitle}{\emph{Proceedings of the Thirteenth International Conference on Artificial Intelligence and Statistics}} \emph{(\bibinfo{series}{Proceedings of Machine Learning Research}, Vol.~\bibinfo{volume}{9})}, \bibfield{editor}{\bibinfo{person}{Yee~Whye Teh} {and} \bibinfo{person}{Mike Titterington}} (Eds.). \bibinfo{publisher}{PMLR}, \bibinfo{address}{Chia Laguna Resort, Sardinia, Italy}, \bibinfo{pages}{249--256}.
\newblock
\urldef\tempurl%
\url{https://proceedings.mlr.press/v9/glorot10a.html}
\showURL{%
\tempurl}


\bibitem[Goodfellow et~al\mbox{.}(2016)]%
        {Goodfellow-et-al-2016}
\bibfield{author}{\bibinfo{person}{Ian Goodfellow}, \bibinfo{person}{Yoshua Bengio}, {and} \bibinfo{person}{Aaron Courville}.} \bibinfo{year}{2016}\natexlab{}.
\newblock \bibinfo{booktitle}{\emph{Deep Learning}}.
\newblock \bibinfo{publisher}{MIT Press}.
\newblock
\newblock
\shownote{\url{http://www.deeplearningbook.org}}.


\bibitem[Hinton and Salakhutdinov(2006)]%
        {hinton2006reducing}
\bibfield{author}{\bibinfo{person}{Geoffrey~E. Hinton} {and} \bibinfo{person}{Ruslan~R. Salakhutdinov}.} \bibinfo{year}{2006}\natexlab{}.
\newblock \showarticletitle{Reducing the dimensionality of data with neural networks}.
\newblock \bibinfo{journal}{\emph{Science}} \bibinfo{volume}{313}, \bibinfo{number}{5786} (\bibinfo{year}{2006}), \bibinfo{pages}{504--507}.
\newblock


\bibitem[Hu et~al\mbox{.}(2024)]%
        {Hu2024}
\bibfield{author}{\bibinfo{person}{Yaofeng Hu}, \bibinfo{person}{Kai Xiao}, \bibinfo{person}{Hengyu Yang}, \bibinfo{person}{Xiaoping Liu}, \bibinfo{person}{Chuanchao Zhang}, {and} \bibinfo{person}{Qianqian Shi}.} \bibinfo{year}{2024}\natexlab{}.
\newblock \showarticletitle{Spatially contrastive variational autoencoder for deciphering tissue heterogeneity from spatially resolved transcriptomics}.
\newblock \bibinfo{journal}{\emph{Briefings in Bioinformatics}} \bibinfo{volume}{25}, \bibinfo{number}{2} (\bibinfo{year}{2024}), \bibinfo{pages}{bbae016}.
\newblock
\showISSN{1477-4054}
\href{https://doi.org/10.1093/bib/bbae016}{doi:\nolinkurl{10.1093/bib/bbae016}}


\bibitem[Jolliffe and Cadima(2016)]%
        {jolliffe2016principal}
\bibfield{author}{\bibinfo{person}{Ian~T. Jolliffe} {and} \bibinfo{person}{Jorge Cadima}.} \bibinfo{year}{2016}\natexlab{}.
\newblock \bibinfo{booktitle}{\emph{Principal Component Analysis} (\bibinfo{edition}{2nd} ed.)}.
\newblock \bibinfo{publisher}{Springer}.
\newblock


\bibitem[Kingma and Welling(2022)]%
        {kingma2022autoencodingvariationalbayes}
\bibfield{author}{\bibinfo{person}{Diederik~P Kingma} {and} \bibinfo{person}{Max Welling}.} \bibinfo{year}{2022}\natexlab{}.
\newblock \bibinfo{title}{Auto-Encoding Variational Bayes}.
\newblock
\showeprint[arxiv]{1312.6114}~[stat.ML]
\urldef\tempurl%
\url{https://arxiv.org/abs/1312.6114}
\showURL{%
\tempurl}


\bibitem[Lee and Seung(1999)]%
        {Lee1999}
\bibfield{author}{\bibinfo{person}{Daniel~D. Lee} {and} \bibinfo{person}{H.~Sebastian Seung}.} \bibinfo{year}{1999}\natexlab{}.
\newblock \showarticletitle{Learning the parts of objects by non-negative matrix factorization}.
\newblock \bibinfo{journal}{\emph{Nature}} \bibinfo{volume}{401}, \bibinfo{number}{6755} (\bibinfo{year}{1999}), \bibinfo{pages}{788--791}.
\newblock
\showISSN{1476-4687}
\href{https://doi.org/10.1038/44565}{doi:\nolinkurl{10.1038/44565}}


\bibitem[Lopez et~al\mbox{.}(2018)]%
        {lopez2018deep}
\bibfield{author}{\bibinfo{person}{Romain Lopez}, \bibinfo{person}{Jeffrey Regier}, \bibinfo{person}{Michael~B Cole}, \bibinfo{person}{Michael~I Jordan}, {and} \bibinfo{person}{Nir Yosef}.} \bibinfo{year}{2018}\natexlab{}.
\newblock \showarticletitle{Deep generative modeling for single-cell transcriptomics}.
\newblock \bibinfo{journal}{\emph{Nature methods}} \bibinfo{volume}{15}, \bibinfo{number}{12} (\bibinfo{year}{2018}), \bibinfo{pages}{1053--1058}.
\newblock


\bibitem[Maas et~al\mbox{.}(2013)]%
        {Maas2013}
\bibfield{author}{\bibinfo{person}{Andrew~L. Maas}, \bibinfo{person}{Awni~Y. Hannun}, {and} \bibinfo{person}{Andrew~Y. Ng}.} \bibinfo{year}{2013}\natexlab{}.
\newblock \showarticletitle{Rectifier Nonlinearities Improve Neural Network Acoustic Models}. In \bibinfo{booktitle}{\emph{Proceedings of the 30th International Conference on Machine Learning (ICML-13).}}
\newblock
\urldef\tempurl%
\url{http://ai.stanford.edu/~amaas/papers/relu_hybrid_icml2013_final.pdf}
\showURL{%
\tempurl}


\bibitem[Rousseeuw(1987)]%
        {Rousseeuw1987}
\bibfield{author}{\bibinfo{person}{Peter~J. Rousseeuw}.} \bibinfo{year}{1987}\natexlab{}.
\newblock \showarticletitle{Silhouettes: A graphical aid to the interpretation and validation of cluster analysis}.
\newblock \bibinfo{journal}{\emph{J. Comput. Appl. Math.}}  \bibinfo{volume}{20} (\bibinfo{year}{1987}), \bibinfo{pages}{53--65}.
\newblock
\showISSN{0377-0427}
\href{https://doi.org/10.1016/0377-0427(87)90125-7}{doi:\nolinkurl{10.1016/0377-0427(87)90125-7}}


\bibitem[Satija et~al\mbox{.}(2015)]%
        {satija2015spatial}
\bibfield{author}{\bibinfo{person}{Rahul Satija}, \bibinfo{person}{Jeffrey~A Farrell}, \bibinfo{person}{David Gennert}, \bibinfo{person}{Alexander~F Schier}, {and} \bibinfo{person}{Aviv Regev}.} \bibinfo{year}{2015}\natexlab{}.
\newblock \showarticletitle{Spatial reconstruction of single-cell gene expression data}.
\newblock \bibinfo{journal}{\emph{Nature biotechnology}} \bibinfo{volume}{33}, \bibinfo{number}{5} (\bibinfo{year}{2015}), \bibinfo{pages}{495--502}.
\newblock


\bibitem[Srivastava et~al\mbox{.}(2014)]%
        {Srivastava2014}
\bibfield{author}{\bibinfo{person}{Nitish Srivastava}, \bibinfo{person}{Geoffrey Hinton}, \bibinfo{person}{Alex Krizhevsky}, \bibinfo{person}{Ilya Sutskever}, {and} \bibinfo{person}{Ruslan Salakhutdinov}.} \bibinfo{year}{2014}\natexlab{}.
\newblock \showarticletitle{Dropout: A Simple Way to Prevent Neural Networks from Overfitting}.
\newblock \bibinfo{journal}{\emph{Journal of Machine Learning Research}}  \bibinfo{volume}{15} (\bibinfo{year}{2014}), \bibinfo{pages}{1929–1958}.
\newblock


\bibitem[Subramanian et~al\mbox{.}(2005)]%
        {Subramanian2005}
\bibfield{author}{\bibinfo{person}{Aravind Subramanian}, \bibinfo{person}{Pablo Tamayo}, \bibinfo{person}{Vamsi~K. Mootha}, \bibinfo{person}{Sayan Mukherjee}, \bibinfo{person}{Benjamin~L. Ebert}, \bibinfo{person}{Michael~A. Gillette}, \bibinfo{person}{Amanda Paulovich}, \bibinfo{person}{Scott~L. Pomeroy}, \bibinfo{person}{Todd~R. Golub}, \bibinfo{person}{Eric~S. Lander}, {and} \bibinfo{person}{Jill~P. Mesirov}.} \bibinfo{year}{2005}\natexlab{}.
\newblock \showarticletitle{Gene set enrichment analysis: A knowledge-based approach for interpreting genome-wide expression profiles}.
\newblock \bibinfo{journal}{\emph{Proceedings of the National Academy of Sciences}} \bibinfo{volume}{102}, \bibinfo{number}{43} (\bibinfo{year}{2005}), \bibinfo{pages}{15545--15550}.
\newblock
\href{https://doi.org/10.1073/pnas.0506580102}{doi:\nolinkurl{10.1073/pnas.0506580102}}
\showeprint{https://www.pnas.org/doi/pdf/10.1073/pnas.0506580102}


\bibitem[Sun et~al\mbox{.}(2024)]%
        {sun2024}
\bibfield{author}{\bibinfo{person}{Yidi Sun}, \bibinfo{person}{Lingling Kong}, \bibinfo{person}{Jiayi Huang}, \bibinfo{person}{Hongyan Deng}, \bibinfo{person}{Xinling Bian}, \bibinfo{person}{Xingfeng Li}, \bibinfo{person}{Feifei Cui}, \bibinfo{person}{Lijun Dou}, \bibinfo{person}{Chen Cao}, \bibinfo{person}{Quan Zou}, {and} \bibinfo{person}{Zilong Zhang}.} \bibinfo{year}{2024}\natexlab{}.
\newblock \showarticletitle{A comprehensive survey of dimensionality reduction and clustering methods for single-cell and spatial transcriptomics data}.
\newblock \bibinfo{journal}{\emph{Briefings in Functional Genomics}} \bibinfo{volume}{23}, \bibinfo{number}{6} (\bibinfo{date}{06} \bibinfo{year}{2024}), \bibinfo{pages}{733--744}.
\newblock
\showISSN{2041-2657}
\href{https://doi.org/10.1093/bfgp/elae023}{doi:\nolinkurl{10.1093/bfgp/elae023}}
\showeprint{https://academic.oup.com/bfg/article-pdf/23/6/733/60974291/elae023.pdf}


\bibitem[Townes and Engelhardt(2023)]%
        {Townes2023}
\bibfield{author}{\bibinfo{person}{F.~William Townes} {and} \bibinfo{person}{Barbara~E. Engelhardt}.} \bibinfo{year}{2023}\natexlab{}.
\newblock \showarticletitle{Nonnegative spatial factorization applied to spatial genomics}.
\newblock \bibinfo{journal}{\emph{Nature Methods}} \bibinfo{volume}{20}, \bibinfo{number}{2} (\bibinfo{year}{2023}), \bibinfo{pages}{229--238}.
\newblock
\showISSN{1548-7105}
\href{https://doi.org/10.1038/s41592-022-01687-w}{doi:\nolinkurl{10.1038/s41592-022-01687-w}}


\bibitem[Traag et~al\mbox{.}(2019)]%
        {Traag2019}
\bibfield{author}{\bibinfo{person}{Vincent~A. Traag}, \bibinfo{person}{Ludo Waltman}, {and} \bibinfo{person}{Nees~Jan van Eck}.} \bibinfo{year}{2019}\natexlab{}.
\newblock \showarticletitle{From Louvain to Leiden: guaranteeing well‐connected communities}.
\newblock \bibinfo{journal}{\emph{Scientific Reports}}  \bibinfo{volume}{9} (\bibinfo{year}{2019}), \bibinfo{pages}{5233}.
\newblock


\bibitem[Wolf et~al\mbox{.}(2018)]%
        {wolf2018scanpy}
\bibfield{author}{\bibinfo{person}{Fabian~Alexander Wolf}, \bibinfo{person}{Philipp Angerer}, {and} \bibinfo{person}{Fabian~J Theis}.} \bibinfo{year}{2018}\natexlab{}.
\newblock \showarticletitle{Scanpy: large-scale single-cell gene expression data analysis}.
\newblock \bibinfo{journal}{\emph{Genome biology}} \bibinfo{volume}{19}, \bibinfo{number}{1} (\bibinfo{year}{2018}), \bibinfo{pages}{1--5}.
\newblock


\bibitem[Wolock et~al\mbox{.}(2019)]%
        {WOLOCK2019281}
\bibfield{author}{\bibinfo{person}{Samuel~L. Wolock}, \bibinfo{person}{Romain Lopez}, {and} \bibinfo{person}{Allon~M. Klein}.} \bibinfo{year}{2019}\natexlab{}.
\newblock \showarticletitle{Scrublet: Computational Identification of Cell Doublets in Single-Cell Transcriptomic Data}.
\newblock \bibinfo{journal}{\emph{Cell Systems}} \bibinfo{volume}{8}, \bibinfo{number}{4} (\bibinfo{year}{2019}), \bibinfo{pages}{281--291.e9}.
\newblock
\showISSN{2405-4712}
\href{https://doi.org/10.1016/j.cels.2018.11.005}{doi:\nolinkurl{10.1016/j.cels.2018.11.005}}


\bibitem[Zhao et~al\mbox{.}(2021)]%
        {Zhao2021}
\bibfield{author}{\bibinfo{person}{Edward Zhao}, \bibinfo{person}{Matthew~R. Stone}, \bibinfo{person}{Xing Ren}, \bibinfo{person}{Jamie Guenthoer}, \bibinfo{person}{Kimberly~S. Smythe}, \bibinfo{person}{Thomas Pulliam}, \bibinfo{person}{Stephen~R. Williams}, \bibinfo{person}{Cedric~R. Uytingco}, \bibinfo{person}{Sarah E.~B. Taylor}, \bibinfo{person}{Paul Nghiem}, \bibinfo{person}{Jason~H. Bielas}, {and} \bibinfo{person}{Raphael Gottardo}.} \bibinfo{year}{2021}\natexlab{}.
\newblock \showarticletitle{Spatial transcriptomics at subspot resolution with BayesSpace}.
\newblock \bibinfo{journal}{\emph{Nature Biotechnology}} \bibinfo{volume}{39}, \bibinfo{number}{11} (\bibinfo{year}{2021}), \bibinfo{pages}{1375--1384}.
\newblock
\showISSN{1546-1696}
\href{https://doi.org/10.1038/s41587-021-00935-2}{doi:\nolinkurl{10.1038/s41587-021-00935-2}}


\end{thebibliography}

\eatme{
\section*{Appendix A: Proof Sketch for Lemma~\ref{lemma:pca}}
In this appendix, we first provide a proof sketch for Lemma~\ref{lemma:pca}, which establishes the equivalence between a linear autoencoder with tied weights and Principal Component Analysis (PCA) in terms of the learned subspace and reconstruction error. This result justifies using PCA as a baseline for linear dimensionality reduction in our benchmarking study of spatial transcriptomics data, where low-dimensional representations of gene expression are critical for tasks like niche discovery and spatial clustering.

\begin{proof}[Proof Sketch]
For a zero-centered data matrix $X \in \mathbb{R}^{n \times d}$, where rows represent $n$ cells and columns represent $d$ genes, PCA projects $X$ onto the subspace of maximum variance. The projection matrix $P_k = U_k U_k^\top$, where $U_k \in \mathbb{R}^{n \times k}$ contains the top-$k$ left singular vectors of $X$, minimizes the reconstruction error $\|X - X P_k\|_F^2$ (Eckart-Young-Mirsky theorem).

A linear autoencoder with tied weights $W \in \mathbb{R}^{k \times d}$ encodes each cell’s gene expression vector $x$ as $Wx$ (a $k$-dimensional latent representation) and decodes it as $W^\top (Wx)$. The reconstructed data is $XWW^\top$, and the autoencoder minimizes:
\[
\min_W \|X - XWW^\top\|_F^2.
\]
Here, $WW^\top$ projects $X$ onto the $k$-dimensional subspace spanned by the columns of $W$, similar to PCA’s projection. As shown in~\cite{baldi1989linear}, the optimal $W^*$ spans the same subspace as the top-$k$ right singular vectors of $X$ (the principal components), because the tied weights ensure an orthogonal projection that maximizes variance.

Thus, the autoencoder’s minimum reconstruction error equals PCA’s: $\|X - XWW^\top\|_F^2 = \|X - X_k\|_F^2$, where $X_k = U_k \Sigma_k V_k^\top$ is the rank-$k$ SVD approximation of $X$. Autoencoders with untied weights or nonlinear activations may achieve different reconstruction errors, depending on the data’s structure and optimization.
\end{proof}

\eatme{
\begin{proof}[Proof Sketch]
When $X$ is centered, the PCA projection matrix $P_k = U_k U_k^\top$ minimizes $\|X - X P_k\|_F^2$, where $U_k$ contains the top $k$ left singular vectors of $X$.

For a linear autoencoder with tied weights $W$, the reconstruction is $XWW^\top$. The objective becomes:
\[
\min_W \|X - XWW^\top\|_F^2.
\]
This is equivalent to projecting $X$ onto the row space of $W$ and then projecting back to the original space via $W^\top$.

It is known (e.g., \cite{baldi1989linear}) that the solution $W^*$ that minimizes this objective spans the same subspace as the top-$k$ PCA directions. Therefore, the optimal reconstruction error of a linear AE matches that of PCA.

Any other AE configuration (e.g., with untied weights or nonlinear activations) may yield higher or lower reconstruction error, depending on the structure of $X$ and optimization dynamics.
\end{proof}
}

\section*{Appendix B: Proof Sketch for Lemma~\ref{lemma:vae}}

This appendix provides a proof sketch for Lemma~\ref{lemma:vae}, which bounds the mutual information between the input $x$ (gene expression profiles) and latent variables $z$ in a Variational Autoencoder (VAE) applied to spatial transcriptomics data. This result, well-known in VAE literature~\cite{Zhao_Song_Ermon_2019}, explains how the KL divergence term regularizes the latent space to capture meaningful patterns like spatial gene expression gradients.

\begin{proof}[Proof Sketch]
By definition, the mutual information between $x$ and $z$ under $q(x, z) = p_{\text{data}}(x) q(z|x)$ is:
\[
I_q(x; z) = \mathbb{E}_{p_{\text{data}}(x)} \left[ \text{KL}(q(z|x) \| q(z)) \right],
\]
where $q(z) = \mathbb{E}_{p_{\text{data}}(x)}[q(z|x)]$ is the aggregated posterior.

We compare $q(z)$ and $p(z)$ using the chain rule of KL divergence:
\[
\text{KL}(q(z|x) \| p(z)) = \text{KL}(q(z|x) \| q(z)) + \text{KL}(q(z) \| p(z)).
\]
Taking expectation over $x \sim p_{\text{data}}(x)$:
\[
\mathbb{E}_{p_{\text{data}}(x)} [\text{KL}(q(z|x) \| p(z))] = I_q(x; z) + \text{KL}(q(z) \| p(z)).
\]
Since $\text{KL}(q(z) \| p(z)) \geq 0$, it follows that:
\[
I_q(x; z) \leq \mathbb{E}_{p_{\text{data}}(x)} [\text{KL}(q(z|x) \| p(z))].
\]
This completes the proof.
\end{proof}

\eatme{
\begin{lemma}
Let $q(z|x)$ be the variational posterior, $p(z)$ the prior, and $q(x, z) = p_{\text{data}}(x) q(z|x)$ the joint distribution. Then the mutual information between $x$ and $z$ satisfies:
\[
I_q(x; z) \leq \mathbb{E}_{p_{\text{data}}(x)} \left[ \text{KL}(q(z|x) \| p(z)) \right].
\]
\end{lemma}

\begin{proof}[Proof Sketch]
By definition, the mutual information between $x$ and $z$ under $q(x, z)$ is:
\[
I_q(x; z) = \mathbb{E}_{p_{\text{data}}(x)} \left[ \text{KL}(q(z|x) \| q(z)) \right],
\]
where $q(z) = \mathbb{E}_{p_{\text{data}}(x)}[q(z|x)]$ is the aggregated posterior.

We now compare $q(z)$ and $p(z)$ via the chain rule of KL divergence:
\[
\text{KL}(q(z|x) \| p(z)) = \text{KL}(q(z|x) \| q(z)) + \text{KL}(q(z) \| p(z)).
\]
Taking expectation over $x$, we get:
\[
\mathbb{E}_{p_{\text{data}}(x)} [\text{KL}(q(z|x) \| p(z))] = I_q(x; z) + \text{KL}(q(z) \| p(z)) \geq I_q(x; z),
\]
since KL divergence is always non-negative.

Thus,
\[
I_q(x; z) \leq \mathbb{E}_{p_{\text{data}}(x)} [\text{KL}(q(z|x) \| p(z))],
\]
which proves the lemma.
\end{proof}
}

\section*{Appendix C: Proof Sketch for Lemma~\ref{lemma:nmf}}

This appendix provides a proof sketch for Lemma~\ref{lemma:nmf}, which establishes the identifiability of Non-negative Matrix Factorization (NMF) under the separability condition. In spatial transcriptomics, NMF decomposes a gene expression matrix into cell contributions and gene modules, and this result ensures that the learned factors are uniquely interpretable, supporting applications like niche discovery~\cite{Arora2012}.

\begin{proof}[Proof Sketch]
For a non-negative data matrix $X \in \mathbb{R}^{n \times d}$, where rows represent $n$ cells and columns represent $d$ genes, NMF seeks to approximate $X \approx WH$, with non-negative $W \in \mathbb{R}^{n \times k}$ (cell contributions) and $H \in \mathbb{R}^{d \times k}$ (gene modules). Under the separability assumption, each column of $H$ has at least one "anchor" row in $X$ that corresponds to a scaled version of that column, isolating its contribution. These anchor points form the vertices of a convex hull in the data space, uniquely determining the columns of $H$. The NMF solution can be constructed by identifying these points, ensuring identifiability up to permutation and scaling of the columns of $W$ and rows of $H$~\cite{Arora2012}.
\end{proof}
\eatme{
\begin{lemma}
Let $X \in \mathbb{R}^{n \times d}$ be a non-negative data matrix. Suppose $X = WH^\top$, with $W \in \mathbb{R}^{n \times k}$, $H \in \mathbb{R}^{d \times k}$, and both $W, H \geq 0$. If $X$ satisfies the separability condition—i.e., each column of $H$ has a row that is an indicator vector—then the factorization is identifiable up to permutation and scaling of the columns of $W$ and $H$.
\end{lemma}

\begin{proof}[Proof Sketch]
Under the separability assumption, every latent component (column of $H$) has at least one "pure" feature — a row in $X$ corresponding to that component only. These anchor points uniquely determine the factorization. The NMF solution can then be constructed by identifying the convex hull of these pure points, which makes the factorization unique up to permutation and scaling \cite{arora2012computing}.
\end{proof}
}

\section*{Appendix D: Proof Sketch for Lemma~\ref{lemma:gae}}

This appendix provides a proof sketch for Lemma~\ref{lemma:gae}, which shows that Graph Autoencoder (GAE) embeddings are smoother over a spatial cell graph than input features in spatial transcriptomics. Smoothness, measured by the trace of the graph Laplacian quadratic form, ensures that nearby cells have similar latent representations, supporting tasks like spatial clustering~\cite{Kipf2017}.

\begin{proof}[Proof Sketch]
For a spatial transcriptomics dataset, let $A$ be the adjacency matrix of a cell graph, where edges reflect spatial proximity, and $L = D - A$ the unnormalized graph Laplacian, with $D$ as the degree matrix. Given input features $X \in \mathbb{R}^{n \times d}$ (gene expression for $n$ cells, $d$ genes), a GAE encoder applies a low-pass spectral filter $g_\theta(L)$, producing embeddings $Z = g_\theta(L)X$. The smoothness functional $\text{Tr}(Z^\top L Z) = \sum_{i,j} A_{ij} \| Z_i - Z_j \|_2^2$ quantifies variation in $Z$ between neighboring cells.

Graph convolutional encoders, such as those based on Graph Convolutional Networks (GCNs), act as low-pass filters by attenuating high-frequency components of the graph spectrum (eigenvalues of $L$). Since high-frequency components correspond to rapid changes across neighbors, $g_\theta(L)$ reduces these variations, ensuring $\text{Tr}(Z^\top L Z) \leq \text{Tr}(X^\top L X)$. This smoothness property holds by construction of the filter~\cite{Kipf2017}.
\end{proof}

\eatme{\begin{lemma}
Let $A$ be the adjacency matrix of a spatial cell graph and $L = D - A$ the unnormalized graph Laplacian. Suppose the encoder of a Graph Autoencoder applies a spectral filter $g_\theta(L)$ to input $X$. Then the learned embedding $Z = g_\theta(L)X$ satisfies:
\[
\text{Smoothness}(Z) = \text{Tr}(Z^\top L Z) \leq \text{Tr}(X^\top L X),
\]
i.e., GAE embeddings are smoother over the graph than the input features.
\end{lemma}

\begin{proof}[Proof Sketch]
Graph convolutional encoders effectively apply a low-pass filter on the graph spectrum. Since $g_\theta(L)$ attenuates high-frequency components, the output $Z$ exhibits reduced variation across connected nodes. The smoothness functional $\text{Tr}(Z^\top L Z)$ quantifies how much $Z$ changes between neighbors. Because $g_\theta(L)$ reduces these fluctuations, smoothness improves by construction.
\end{proof}
}

\section*{Appendix E: Proof Sketch for Lemma~\ref{lemma:snmf}}

This appendix provides a proof sketch for Lemma~\ref{lemma:snmf}, which introduces a spatially-regularized Non-negative Matrix Factorization (NMF) for spatial transcriptomics. By combining NMF’s non-negativity with a graph Laplacian smoothness constraint, the lemma ensures that cell embeddings are both interpretable and spatially coherent, enhancing applications like niche discovery~\cite{Arora2012}.

\begin{proof}[Proof Sketch]
Let $X \in \mathbb{R}^{n \times d}$ be a non-negative data matrix (cells by genes), $A$ the adjacency matrix of a spatial cell graph, and $L = D - A$ the unnormalized graph Laplacian. The optimization problem is:
\[
\min_{W,H \geq 0} \|X - WH\|_F^2 + \lambda \text{Tr}(W^\top L W),
\]
where $W \in \mathbb{R}^{n \times k}$ represents cell embeddings, $H \in \mathbb{R}^{d \times k}$ represents gene modules, and $\lambda > 0$ controls smoothness.

The smoothness term $\text{Tr}(W^\top L W) = \sum_{i,j} A_{ij} \| W_i - W_j \|_2^2$ penalizes differences in embeddings between neighboring cells. Since $L$ is positive semi-definite, $\text{Tr}(W^\top L W) \geq 0$, and the optimization encourages $W$ to have small differences across edges, reducing $\text{Tr}(W^\top L W)$ compared to $\text{Tr}(X^\top L X)$, where $X$ lacks this constraint.

To show $\text{Tr}(W^{*\top} L W^*) \leq \text{Tr}(X^\top L X)$, consider the optimization dynamics. The gradient of the smoothness term encourages $W$ to align with the low-frequency eigenvectors of $L$, similar to graph convolutional filters (cf. Lemma~\ref{lemma:gae}). Since $X$ is not optimized for smoothness, its variation across neighbors is typically higher, implying $\text{Tr}(W^{*\top} L W^*) \leq \text{Tr}(X^\top L X)$.

For identifiability, assume $X$ satisfies the separability condition (cf. Lemma~\ref{lemma:nmf}), where each column of $H$ has an anchor row in $X$. The smoothness term does not affect the convex hull of anchor points, as it operates on $W$. Thus, the factorization remains identifiable up to permutation and scaling, as shown in \cite{Arora2012}.

This proves the lemma.
\end{proof}
}
\end{document}